\documentclass[12pt]{article}
\usepackage{times}
\usepackage[utf8]{inputenc}
\usepackage{setspace}
\usepackage[margin=1.0in]{geometry} 
\usepackage{enumitem} 
\usepackage{xcolor} 
\usepackage{soul} 

\usepackage{mathtools}
\usepackage{bm}
\usepackage{amsmath} 
\usepackage{dsfont} 
\usepackage{cancel}
\usepackage[normalem]{ulem} 
\newcommand{\stkout}[1]{\ifmmode\text{\sout{\ensuremath{#1}}}\else\sout{#1}\fi} 
\DeclareMathOperator*{\argmax}{argmax}

\usepackage{amssymb} 

\usepackage{graphicx}
\graphicspath{{images/}}

\usepackage{float} 
\usepackage{adjustbox}
\usepackage{caption}

\usepackage{algorithm2e} 
\RestyleAlgo{ruled} 

\usepackage{datetime}
\newdateformat{monthdayyeardate}{%
  \monthname[\THEMONTH]~\THEDAY, \THEYEAR}
\usepackage{authblk} 
\providecommand{\keywords}[1] 
{
  \small	
  \textbf{Keywords } #1 
}

\usepackage{hyperref} 
\hypersetup{
	colorlinks=true,
    linkcolor=blue,
    citecolor=blue,
    filecolor=magenta,      
    urlcolor=blue 
}
\title{A new algorithm for sampling parameters in a structured correlation matrix with application to estimating optimal combinations of muscles to quantify progression in Duchenne muscular dystrophy}
\author[1]{Michael K. Kim}
\author[1*]{Michael J. Daniels}
\author[2]{William D. Rooney}
\author[3]{Rebecca J. Willcocks}
\author[4]{Glenn A. Walter}
\author[3]{Krista H. Vandenborne}
\affil[1]{Department of Statistics, University of Florida, Gainesville, Florida, USA}
\affil[2]{Advanced Imaging Research Center, Oregon Health and Science University, Portland, Oregon, USA}
\affil[3]{Department of Physical Therapy, University of Florida, Gainesville, Florida, USA}
\affil[4]{Department of Physiology and Aging, University of Florida, Gainesville, Florida, USA}
\affil[*]{\textbf{Correspondence}: Michael J. Daniels, Department of Statistics, University of Florida, 230 Newell Drive. Gainesville, FL, 32611, USA. Email: \url{daniels@ufl.edu}}

\date{\monthdayyeardate\today}

\begin{document}

\maketitle

	\begin{abstract}
	\noindent
	The goal of this paper is to estimate an optimal combination of biomarkers for individuals with Duchenne muscular dystrophy (DMD), which provides the most sensitive combinations of biomarkers to assess disease progression (in this case, optimal with respect to standardized response mean (SRM) for 4 muscle biomarkers). The biomarker data is an incomplete (missing and irregular) multivariate longitudinal data. We propose a normal model with structured covariance designed for our setting. To sample from the posterior distribution of parameters, we develop a Markov Chain Monte Carlo (MCMC) algorithm to address the positive definiteness constraint on the structured correlation matrix. In particular, we propose a novel approach to compute the support of the parameters in the structured correlation matrix; we modify the approach from \cite{Barnard} on the set of largest possible submatrices of the correlation matrix, where the correlation parameter is a unique element. For each posterior sample, we compute the optimal weights of our construct. We conduct data analysis and simulation studies to evaluate the algorithm and the frequentist properties of the posteriors of correlations and weights. We found that the lower extremities are the most responsive muscles at the early and late ambulatory disease stages and the biceps brachii is the most responsive at the non-ambulatory disease stage.
	\end{abstract}
	
	
	\section{Introduction} \label{Introduction}
	For Bayesian inference using multivariate normal (MVN) models with a structured correlation matrix, posterior computation can be challenging due to positive definiteness restrictions. We introduce an approach based on \cite{Barnard} to address this.
	\\ \\
	We use this model to address an important problem in Duchenne muscular dystrophy (DMD) -- finding an optimal combination of biomarkers. The data are annualized changes of fat fraction (FF) of muscles (which are the biomarkers). We want to compute an optimal combination of these biomarkers to more precisely evaluate disease progression and potentially have more sensitive endpoints for clinical trials. The data available to do this is multivariate longitudinal at irregular time points with substantial missingness. We will use a multivariate normal model with structured correlation to model this data and address the missingness. Previous unpublished work only allowed for complete data.
	\\ \\
	We first review relevant literature on covariance/correlation matrices. To ensure positive definiteness of the covariance matrix, multiple works have relied on unconstrained parameterizations of either the covariance or correlation matrix. Some earlier works regarding the covariance matrix include the Cholesky decomposition \cite{Lindstrom}, the matrix logarithm \cite{Leonard} \cite{Chiu}, or Givens angles \cite{Yang} \cite{DanielsKass1999}. These techniques focus on estimation and do not allow interpretation of new parameters in terms of original variances and correlations. Later works try to address this issue, and one of the most popular examples is the modified Cholesky decomposition \cite{Pourahmadi1999}. Their utility is limited by the ordering of variables which is typically only natural for univariate longitudinal data. Other papers that exploit the ordering are \cite{Pan2003}, \cite{Pan2006}, \cite{Leng2010}, \cite{ZhangLeng2012}.
	\\ \\
	There have also been works for the unconstrained parameterization methods for the correlation matrix, which are of interest in Bayesian settings that deal with correlations and variances separately. A correlation matrix has the additional constraint of diagonal elements fixed at one. Pinheiro and Bates \cite{Pinheiro} introduce the spherical parameterization, which is conducted by first performing the Cholesky decomposition on the correlation matrix and then expressing the elements of the Cholesky factor as angles. More recent works expanded on this \cite{Zhang} \cite{Tsay}. Due to the Cholesky decomposition, the angles are dependent on ordering. The angles are defined on the support $(0,\pi]$ to ensure uniqueness and identifiability. Namely, there is a monotone relationship between the angles and the correlations -- weaker correlations imply larger angles. For longitudinal data, this implies angles are increasing with time lag.
	\\ \\
	Zhang et al. \cite{Zhang} illustrate the applicability of the spherical parameterization for unbalanced longitudinal data by proposing a joint mean-variance-correlation generalized linear model. As spherical parameterization is dependent on ordering of variables, their model only permits popular correlation structures for longitudinal data such as compound symmetry or AR(1). This is problematic for complex correlated data with partial ordering information, including longitudinal data that has multiple outcomes measured repeatedly from the same subject (i.e. multivariate longitudinal data). Tsay and Pourahmadi \cite{Tsay} address this by showing that positive definiteness is guaranteed for structured correlation matrices by using spherical parameterization alongside pivotal angles. By knowing the locations and estimates of pivotal angles, one can obtain the unique correlations and then the implied (``non-pivotal") angles row-by-row. Both papers use maximum likelihood estimation as it achieves consistency and asymptotic normality. However, if were to apply Bayesian computation on the angles, we would get different posteriors based on how the data was ordered, as typical priors on the angles are not invariant to ordering. Ghosh et al. \cite{Ghosh} introduce shrinkage and selection priors for the angles which are the inverse cosine of semi-partial correlations, but the priors are only applicable for ordered data. Similar to \cite{Zhang}, Ghosh et al. \cite{Ghosh} only consider a limited set of correlation structures: AR(1), banded, and block common.
	\\ \\
	Another unconstrained parameterization method for the correlation matrix uses the partial autocorrelations adjusted for intervening variables \cite{Daniels2009} \cite{Wang2013}. Similar to the angles of hyperspherical parameterization, partial autocorrelations impose an order on the variables and have a one-to-one and recursive relationship with the correlations. This results in similar interpretability issues as before, but the authors do introduce priors for the correlation matrix that are independent of the order of indices. For application, they explore parsimonious modeling for balanced longitudinal (ordered) data with a focus on lags. More recent works take a different approach and apply the matrix logarithm on the correlation matrix \cite{Archakov} \cite{Hu}. The transformation is one-to-one, invariant to ordering, and offers flexibility for parsimonious modeling and prior specification of unbalanced data. However, interpretability of the new parameters is not intuitive.
	\\ \\
	A notable constrained approach for the covariance matrix is the linear covariance model \cite{Anderson}, where the covariance matrix is the linear combination of known symmetric matrices and unknown coefficients. Constraints are put on the unknown coefficients, and the approach is applicable for any element-wise estimation of the covariance matrix. On the other hand, a constrained approach that directly models the correlation elements avoid issues with both interpretability and the unity diagonal constraint of the correlation matrix, both posing challenges for many unconstrained parameterization methods. The constrained Bayesian approach puts the positive definiteness constraint in each sampling step. Some earlier works are \cite{Barnard}, \cite{Wong}, \cite{Liechty}. Barnard et al. \cite{Barnard} note that a correlation matrix stays positive definite if one were to replace any unique correlation element inside of it from an interval calculated using the values of all other correlation elements. They demonstrate the effectiveness of this approach with order-invariant priors. They use independent log normal priors for standard deviations and either the marginally uniform priors for correlations or the jointly uniform prior for the correlation matrix. Wong et al. \cite{Wong} estimate the covariance matrix of normal data by identifying zeroes in its inverse (see covariance selection models \cite{Dempster}), separating the inverse into a product of inverse partial variances and the matrix of partial correlations, and using `covariance selection' priors that allows zeroes in the precision matrix. There is a constraint on the prior for the matrix of partial correlations. The efficiency of their method and any identification of structure rely on the sparsity of the true precision matrix. Pitt et al. \cite{Pitt} is an extension and Carter et al. \cite{Carter} is a generalization of \cite{Wong}.
	\\ \\
	Similar to \cite{Barnard}, Liechty et al. \cite{Liechty} puts the positive definiteness constraint on the prior of the correlation matrix and compute intervals to sample correlation elements. They propose prior probability models that group marginal correlations into clusters based on similarities among correlations or variables. This results in group-structured correlation matrices, i.e. block common. The `common correlation' prior allows shrinkage towards a diagonal correlation matrix. The `grouped correlation' and `grouped variables' mixture priors have flexibility in shrinkage towards target structures by using a point mass at zero or a small-variance distribution for a term in the mixture. Zhang et al. \cite{Zhang2022} is a recent work in constrained Bayesian approach for unstructured and unordered correlation matrices that are functions of individual-level covariates, where each correlation element is specified by a linear model. They focus on how positive definiteness is ensured at different values of covariates. They intersect variations of the intervals defined in \cite{Barnard} to address the positive definiteness constraint of the correlation matrices. We also intersect `Barnard' intervals but based on submatrices in the setting of a structured correlation matrix.
	\\ \\
	We provide a flexible approach to modeling correlation matrices by generalizing the interval approach of \cite{Barnard} for a structured correlation matrix. In our application, it is a correlation matrix of multiple outcomes measured at each time point in an unbalanced longitudinal data with missingness. Our structure for the application assumes exchangeable time points, but the approach can be used to model any correlation structure without requiring an ordering of variables.
	\\ \\
	The paper is organized as follows. In Section \ref{Data and motivation}, we introduce the DMD data and explain the clinical motivation of an optimal combination of biomarkers. In Section \ref{Model and objective function}, we introduce the model and the structure of the correlation matrix. We also introduce the objective function to optimally choose weights of the construct. In Section \ref{Posterior computation}, we provide details on the posterior computation. In particular, details on how to compute a `tight' positive definite (PD) interval of the candidate distribution of a correlation element by applying the interval approach of \cite{Barnard} on the set of the largest possible submatrices of the correlation matrix. In the simulation study of Section \ref{Simulations}, we compare the acceptance and positive definiteness rates of correlations drawn from candidate distributions that use our generated PD intervals vs. $(-1,1)$ support. Section \ref{Simulations} also compares the computational performance and frequentist operating characteristics via simulations. Section \ref{Data analysis} presents our findings on the DMD data, both clinically and algorithmically. Section \ref{Discussion} provides conclusions and extensions.
	
	\section{Data and motivation} \label{Data and motivation}
	The muscles of individuals with Duchenne muscular dystrophy (DMD) are progressively replaced by fat but at different rates, and these rates vary by individual, age, disease stage, and other factors. The disease stages we consider are defined by functional ability. In the early ambulatory stage, individuals can walk and get up from the floor without assistance. In the late ambulatory stage, individuals can walk but can no longer get up from the floor. In the non-ambulatory stage, individuals cannot walk. The muscles of interest are two lower extremity muscles, soleus (SOL) and vastus lateralis (VL), and two upper extremity muscles, biceps brachii (BB) and deltoid (DEL). We use magnetic resonance spectroscopy fat fraction (FF) measures of the muscles \cite{Forbes}. The specific measurement we use is the annualized change in FF from the current and previous visit of a subject, where visits are somewhere between 6--24 months apart. It is calculated as $(\text{FF at current visit} - \text{FF at previous visit})/(\text{age at current visit} - \text{age at previous visit})$.
	For brevity, we use ``muscle" to describe FF changes in a muscle between visits. Also, we use ``time point" interchangeably with ``measurement time."
	\\ \\
	The numbers of subjects (160 unique subjects) and visits vary by ambulatory disease stage. In total, early ambulatory data has 140 subjects and 419 measurement times, late ambulatory data has 74 subjects and 128 measurement times, and non-ambulatory data has 51 subjects and 115 measurement times. The data exhibits missingness, as not all muscles are measured for each measurement time. Tables \ref{table:DM1}, \ref{table:DM2}, and \ref{table:DM3} in the \hyperref[Supplement A]{Supporting Information} provides details: Table \ref{table:DM1} provides the missingness by muscle, revealing significantly more missinginess in the upper extremities than the lower extremities. Table \ref{table:DM2} provides the distribution of the number of missing muscle measurements at a given measurement time. Typically two muscles are missing at a given measurement time, particularly in earlier ambulatory stages. Table \ref{table:DM3} provides the distribution of the number of subject measurement times for the unbalanced DMD data. There is a maximum of 8, 6, and 7 subject measurement times for early, late, and non-ambulatory data, respectively.
	\\ \\
	To account for the heterogeneity in the fat replacement rate of different muscles, we want a sensitive measure of overall muscle quality across different disease stages. We will construct an optimal combination of biomarkers that produces the most clinically meaningful and sensitive combination of FF measures of different muscles with varying weighting coefficients across a wide spectrum of disease stages. The inclusion of non-ambulant subjects here is important. The majority of individuals with DMD are non-ambulatory but are excluded from most clinical trials due to the inapplicability of traditional functional outcome measures \cite{Verhaart}.
	
	\section{Model and objective function} \label{Model and objective function}
	\subsection{Model} \label{Model}
	For an individual with muscular dystrophy in a certain ambulatory disease stage, we denote the data as $y_{ij\ell}$, indicating the FF of subject $i$, measurement time $j$ and muscle $\ell$. In total, we have $N$ subjects, $J$ measurement times, and $L$ muscles. Let $p=J \times L$ be the total number of outcomes for a subject. The model below is for unbalanced longitudinal data, but for simplicity of notation, we will use $J$ and $p$ rather than $J_{i}$ and $p_{i}$.
	\\ \\
	We model the data by a multivariate normal distribution and assume exchangeable time points for a subject in each ambulatory disease stage. We employ the separation strategy on the covariance matrix \cite{Barnard}:
	$$\bm{Y}_{i} \stackrel{\text{ind}}{\sim}\text{MVN}(\bm{\mu}, \bm{\Sigma})$$
	$$\bm{\Sigma}=\bm{S}\bm{R}\bm{S}.$$
	Due to the assumption of exchangeable time points, we put structure on the mean vector, the diagonal standard deviation matrix, and the correlation matrix as follows,
	\begin{itemize}[noitemsep]
		\item $\bm{\mu}_{p \times 1}=(\tilde{\bm{\mu}}'_{L \times 1},...,\tilde{\bm{\mu}}_{L \times 1}')'$ is a mean vector with unique elements $\tilde{\bm{\mu}}=(\mu_{1},...,\mu_{L})'$ for each measurement time $j$, where $\mu_{\ell}$ is the mean of muscle $\ell$, i.e. $\text{E}[Y_{ij\ell}]=\mu_{\ell}$.
		\item $\bm{S}_{p \times p}=\text{diag}(\bm{s}'_{L \times 1}, ..., \bm{s}'_{L \times 1})$ is a diagonal standard deviation matrix with unique elements $\bm{s}=(s_{1},...,s_{L})'$ for each measurement time $j$, where $s_{\ell}$ is the standard deviation of muscle $\ell$, i.e. $\text{SD}[Y_{ij\ell}]=s_{\ell}$.
		\item $\bm{R}_{p \times p}$ is a structured correlation matrix with unique elements 
		$$\bm{r}=(\underbrace{\eta_{12},...,\eta_{L-1,L}}_{{L \choose 2}}, \underbrace{\rho_{(1)},...,\rho_{(L)}}_{L}, \gamma)'=(r_{1},...,r_{q})', \ \text{where}$$
		\begin{itemize}[nosep]
			\item $\eta_{\ell\ell'}$ is the correlation between two different muscles $\ell$ and $\ell'$ observed at the same measurement time, i.e. $\text{Corr}[Y_{ij\ell}, Y_{ij\ell'}]=\eta_{\ell\ell'}$. There are ${L \choose 2}$ $\eta_{\ell\ell'}$'s, and for each $\eta_{\ell\ell'}$, there are $J$ instances of it in the upper-triangular portion of $\bm{R}$, so in total there are $JL(L-1)$ $\eta_{\ell\ell'}$'s in $\bm{R}$.
			\item $\rho_{(\ell)}$ is the correlation between any two different measurement times for muscle $\ell$, i.e. $\text{Corr}[Y_{ij\ell}, Y_{ij'\ell}]=\rho_{(\ell)}$. There are $L$ $\rho_{(\ell)}$'s, and for each $\rho_{(\ell)}$, there are ${J \choose 2}$ instances of it in the upper-triangular portion of $\bm{R}$, so in total there are $JL(J-1)$ $\rho_{(\ell)}$'s in $\bm{R}$.
			\item $\gamma$ is the correlation between two different muscles at two different measurement times, i.e. \\
			$\text{Corr}[Y_{ij\ell}, Y_{ij'\ell'}]=\gamma$. There are $L^2J^2-2J{L \choose 2}-2L{J \choose 2}-LJ$ or $JL(J-1)(L-1)$ $\gamma$'s in $\bm{R}$.
			\item The matrix has $q={L \choose 2}+L+1=11$ unique parameters.
		\end{itemize}
	\end{itemize}
	The structured correlation matrix for $L=4$ muscles is as follows:
	$$\bm{R}=\begin{bsmallmatrix}
			1 & \eta_{12} & \eta_{13} & \eta_{14} & \rho_{(1)} & \gamma & \gamma & \gamma & \rho_{(1)} & \gamma & \gamma & \gamma & ... \\
			& 1 & \eta_{23} & \eta_{24} & \gamma & \rho_{(2)} & \gamma & \gamma & \gamma & \rho_{(2)} & \gamma & \gamma & ... \\
			& & 1 & \eta_{34} & \gamma & \gamma & \rho_{(3)} & \gamma & \gamma & \gamma & \rho_{(3)} & \gamma & ... \\
			& & & 1 & \gamma & \gamma & \gamma & \rho_{(4)} & \gamma & \gamma & \gamma & \rho_{(4)} & ... \\
			& & & & 1 & \eta_{12} & \eta_{13} & \eta_{14} & \rho_{(1)} & \gamma & \gamma & \gamma & ...\\
			& & & & & 1 & \eta_{23} & \eta_{24} & \gamma & \rho_{(2)} & \gamma & \gamma & ...\\
			& & & & & & 1 & \eta_{34} & \gamma & \gamma & \rho_{(3)} & \gamma & ... \\
			& & & & & & & 1 & \gamma & \gamma & \gamma & \rho_{(4)} & ... \\
			& & & & & & & & 1 & \eta_{12} & \eta_{13} & \eta_{14} & ... \\
			& & & & & & & & & 1 & \eta_{23} & \eta_{24} & ... \\
			& & & & & & & & & & 1 & \eta_{34} & ... \\
			& & & & & & & & & & & 1 & ... \\
			\vdots \\
		\end{bsmallmatrix}.$$
 	Given the objective of accounting for correlation within disease stage windows, we assume exchangeable time points for simplicity. But we could easily accommodate time series structures including autoregressive structures.
       
	\subsection{Optimal construct} \label{Optima l construct}
	For an individual with muscular dystrophy at a certain ambulatory disease stage, we compute an optimal convex combination $\tilde{y}_{ij}=\sum_{\ell=1}^{L}w_{\ell}y_{ij\ell}=\bm{w}'\bm{y}_{ij}$, where $\sum_{\ell=1}^{L}w_{\ell}=1$, $w_{\ell} \geq 0 \ \forall \ell$. The weights $\bm{w}$ are chosen to maximize the standardized response mean (SRM) of the construct, where $\bm{\Sigma}_{\eta}=\text{Var}[\bm{Y}_{ij}]$ is the $L \times L$ covariance matrix containing the correlation parameters, $\eta_{\ell\ell'}$'s:
	\begin{equation} \label{eq:w}
	\bm{w}=\argmax_{\bm{w}}\text{SRM}[\tilde{Y}_{ij}]=\argmax_{\bm{w}}\frac{\text{E}[\tilde{Y}_{ij}]}{\text{SD}[\tilde{Y}_{ij}]}=\argmax_{\bm{w}}\frac{\bm{w}'\tilde{\bm{\mu}}}{\sqrt{\bm{w}'\bm{\Sigma}_{\eta}\bm{w}}}.
	\end{equation}
	We note that the standard deviation of the construct does not vary across time, but the weights are optimized for each ambulatory disease stage. These weights provide the largest mean of convex combinations of 1-year changes in FF of muscles with respect to variability in the changes for each disease stage.
	
	\section{Posterior computation} \label{Posterior computation}
	Full details of the Markov Chain Monte Carlo (MCMC) algorithm for posterior computation are presented in Section \ref{Supplement - Details of MCMC algorithm} of the \hyperref[Supplement A]{Supporting Information}. Here we focus on our new approach for sampling the correlations and computing the optimal biomarker constructs.
	
	\subsection{Sampling the structured correlation parameters} \label{Sampling correlations}
	In a Metropolis-Hastings (M-H) step, when the correlation matrix has dimension greater than 2, a correlation candidate may produce a non-PD correlation matrix if we were to sample from a candidate distribution with a $(-1,1)$ support.
	To mitigate this, the support of the candidate distribution should be shrunk to a tighter interval.
	Barnard et al. \cite{Barnard} introduced an approach that computes an interval for a unique correlation element inside of a correlation matrix by solving a quadratic equation using determinants of augmented correlation matrices (details in Section \ref{Supplement - Barnard approach details} of the \hyperref[Supplement A]{Supporting Information}). The correlation matrix remains positive definite as long as the value of the correlation element lies within the interval. Their approach works for any element inside of an unstructured correlation matrix. We will refer to their approach as the ``Barnard approach" and the interval computed from the approach as the ``PD interval."
	\\ \\
	We adapt the Barnard approach on a structured correlation matrix $\bm{R}$ with less unique elements $(r_{1},...,r_{q})'$ than an unstructured correlation matrix. We first create the set of all the largest possible submatrices of $\bm{R}$, where each of these submatrices contains the $k$th correlation $r_{k} \in (r_{1},...,r_{q})'$ only once, and the rest of the submatrix is filled with current values of correlation elements other than the $k$th correlation $\bm{r}_{(-k)}^{(t-1)}$.
	After computing the PD interval for each submatrix, we define the support of the candidate distribution of $r_{k}^{(t)}$ as the intersection of all the PD intervals. If there is only one largest possible submatrix, then we use its PD interval as  the support. We denote the support as $(L_{k}^{(t)}, U_{k}^{(t)})$. 
	Our approach improves the rate at which a correlation candidate results in a PD $\bm{R}$.
	\\ \\
	To find the largest possible submatrix, the elements of the correlation matrix need to be reordered. In particular, we can build a largest possible submatrix for $r_{k}$ with respect to \textit{any} structured correlation matrix by selecting a subset of the original outcomes $\hat{\bm{y}} \subset \bm{y}=(y_{1},...,y_{p})'$ using Algorithm \ref{general_algo} in the \hyperref[Supplement A]{Supporting Information}, which is summarized below:
	\begin{enumerate}[noitemsep]
		\item The first two outcomes in the subset should correspond to correlation $r_{k}$.
		\item Under the condition that $r_{k}$ remains as a unique element inside the submatrix, check one-by-one whether the remaining outcomes can be added to the subset
	\end{enumerate}
	To compute the tightest (intersected) PD interval, we use all combinations of the largest possible submatrix for $r_{k}$. We can apply Algorithm \ref{general_algo} in the \hyperref[Supplement A]{Supporting Information} on all permutations of the ordering of outcomes. This can be computationally expensive and can result in duplicate PD intervals.
	\\ \\
	To save computational time for our structure specified in Section \ref{Model}, we devised individualized algorithms for $r_{1},...,r_{q}$, efficiently generating all combinations of the largest possible submatrix for each $r_{k}$ (Section \ref{Supplement - Building the largest} of the \hyperref[Supplement A]{Supporting Information}). An alternative to further reduce computational cost is to compute the PD interval of only one submatrix for each $r_{k}$, particularly when the submatrix has dimension close to that of the correlation matrix.  We explore this option in Section~\ref{Simulations}.
	\\ \\
	For the candidate distribution, we recommend two choices. First, a uniform distribution on the derived PD interval,
	\begin{equation} \label{eq:Unif(L,U)}
	r_{k}^{(t)} \sim \text{Unif}(L_{k}^{(t)}, U_{k}^{(t)}). 
	\end{equation}
	Second, a reparameterized-beta with mode at the value of previous iteration,
	\begin{equation} \label{eq:R-Beta(L,U)}
	r_{k}^{(t)} \sim \text{R-Beta}(\alpha_{k}^{(t)}, \beta_{k}^{(t)}, L_{k}^{(t)}, U_{k}^{(t)}), \text{ where}
	\end{equation}
	$$\alpha^{(t)}_{k}=\frac{(\kappa_{k}-1)L_{k}^{(t)}+(2-\kappa_{k})r_{k}^{(t-1)}-U_{k}^{(t)}}{L_{k}^{(t)}-U_{k}^{(t)}},$$
	$$\beta^{(t)}_{k}=\frac{(\kappa_{k}-1)U_{k}^{(t)}+(2-\kappa_{k})r_{k}^{(t-1)}-L_{k}^{(t)}}{U_{k}^{(t)}-L_{k}^{(t)}}=\kappa_{k}-\alpha^{(t)}_{k},$$
	$\kappa_{k}$ is the concentration (and tuning) parameter, and the mode is $r_{k}^{(t-1)}$, which is constrained to the interval $(L_{k}^{(t)}, U_{k}^{(t)})$.
	\\ \\
	If the reparameterized-beta distribution is used, we can tighten the interval we draw $r_{k}^{(t)}$ from by setting $\kappa_{k}$ to a higher value, which makes the candidate distribution be more centered around the previous value $r_{k}^{(t-1)}$. Doing this will increase the percentage of times that $\bm{R}$ is PD throughout the M-H for $r_{k}$, since we would be reducing the possible candidates that would be far away from $r_{k}^{(t-1)}$ in which it is ensured to produce a PD $\bm{R}$. In turn, the acceptance rate for $r_{k}$ will also increase, since the prior forces the M-H algorithm to automatically reject an $r_{k}^{(t)}$ that would produce a non-PD $\bm{R}$. Candidate distributions like the reparameterized-beta distribution allow the researcher to achieve the standard 25\% M-H acceptance rate for $r_{k}$ by adjusting $\kappa_{k}$ \cite{Gelman}. However, there will be more potential autocorrelation in a chain given the centering on the previous iteration value.
    Candidate distributions like the uniform distribution allow for a more even exploration of correlation candidates on $(L_{k}^{(t)}, U_{k}^{(t)})$ that would not be directly influenced by $r_{k}^{(t-1)}$.
	
	\subsection{Computing the posterior distribution of the weights} \label{Computing the posterior distribution of the weights}
	The standardized response mean (SRM) of $\tilde{Y}_{ij}$ is a function of parameters $\tilde{\bm{\mu}}$ and $\bm{\Sigma}_{\eta}$. Therefore, for each ambulatory disease stage and each posterior sample of $\tilde{\bm{\mu}}$ and $\bm{\Sigma}_{\eta}$, we can compute the value of $\bm{w}$ that maximizes the SRM (Equation \ref{eq:w}). The posterior point estimate of the weights has to account for the simplex constraints, so we use the set of weights that optimize the SRM, by plugging in the posterior medians of the model parameters defined in Section \ref{Model}. Computing a point estimate of each weight directly does not guarantee that the weights sum to 1, and given the slight skewness and unimodal shape of the posterior distributions of all model parameters, the median was chosen as a robust point estimate.
	
	\section{Simulations} \label{Simulations}
	We conduct simulations to assess frequentist operating characteristics of the optimal weights and SRM. In particular, we evaluate the coverages of 95\% credible intervals, biases, and mean squared errors (MSEs) of the weights and SRM by using simulated data.
	\\ \\
	We also do simulations to examine the acceptance and positive definiteness rates of correlations drawn using our proposed algorithm.
	\\ \\
	To generate the simulated data for each ambulatory disease stage, we use the posterior medians of the model parameters for the DMD data (Table \ref{table:DA10} in the \hyperref[Supplement A]{Supporting Information}) as the truth. We generate 500 simulated datasets for $N=100$ subjects and $J=4$ time points. We consider four true distributions: normal (Section \ref{Model}), $t_{10}$ \cite{Kotz}, $t_{3}$, and skew-normal with shape (skewness) parameter of $0.1$ \cite{Azzalini1999}. We also generate simulated data with missingness using the normal model. We emulate the column-wise missingness and row-wise missingness patterns of the DMD data (Tables \ref{table:DM1} and \ref{table:DM2} in the \hyperref[Supplement A]{Supporting Information}) by using $(0.05, 0.05, 0.75, 0.75)'$ and $(0.1, 0.6, 0.1, 0.2)'$, respectively.
	\\ \\
	For each simulated dataset, we run 50000 iterations with 1000 burn-in. Unless otherwise noted, posterior results assume that correlations were sampled using a Metropolis-Hastings (M-H) algorithm with candidate distribution, reparameterized-beta (Equation \ref{eq:R-Beta(L,U)}) on the PD interval described in Section \ref{Sampling correlations}. We will refer to this distribution as $\text{R-Beta}(L,U)$. We will refer to Equation \ref{eq:Unif(L,U)} as $\text{Unif}(L,U)$.
	 
	\subsection{Coverages, biases, and MSEs of weights and SRM} \label{Coverage, bias, MSE of weights and SRM}
	Coverages of 95\% credible intervals of weights and SRM are given in Table \ref{table:sim1}. The coverages for normal (with missingness) are close to those for normal. For $t_{10}$, coverages of weights are close to those for normal, but there is some undercoverage with SRM.  For $t_{3}$, there is considerable undercoverage with the weights due to the distribution's heavy tails, but there is overcoverage for true weights equal to zero. For skew-normal, there is mostly undercoverage with the weights and severe undercoverage with SRM.
    \\ \\
    Biases of weights and SRM are given in Table \ref{table:sim2}. The magnitudes of the biases are relatively small with respect to the weights $\in (0,1)$ and SRM $\in (0.8,1.4)$. Biases of weights slightly increase from normal to $t_{10}$, moderately increase from $t_{10}$ to $t_{3}$, and considerably increase from $t_{3}$ to skew-normal. Biases of SRM slightly increases from normal to $t_{10}$, then considerably increases from $t_{10}$ to $t_{3}$ and from $t_{3}$ to skew-normal.
    \\ \\
    Root MSEs of weights and SRM are given in Table \ref{table:sim3_extra} in the \hyperref[Supplement A]{Supporting Information}. Root MSEs of weights generally have moderate values for normal, $t_{10}$, and skew-normal. MSEs of weights for $t_{3}$ and MSEs of SRM for $t_{3}$ and skew-normal are large. Some of the smallest MSEs are for true weights equal to zero. It is expected that biases and MSEs for simulated data with missingness are slightly larger than those for simulated data without missingness. For SRM, there is a noticeable decrease in coverage and increase in bias and MSE from $t_{10}$ to $t_{3}$ and from $t_{3}$ to skew-normal. The poorer frequentist operating characteristics are expected for a mis-specified observed data model.

	\subsection{Acceptance and positive definiteness rates of drawn correlations} \label{Acceptance and positive definiteness rates of correlations}

	This study is conducted with the normal distribution scenario and no missingness. It is meant to evaluate the proposed approach for sampling correlation parameters of the structured correlation matrix specified in Section \ref{Model}. We compare the M-H acceptance and positive definiteness rates of correlations drawn from $\text{R-Beta}(L,U)$, $\text{R-Beta}(L_{1},U_{1})$, $\text{R-Beta}(-1,1)$, $\text{Unif}(L,U)$, $\text{Unif}(L_{1},U_{1})$, and $\text{Unif}(-1,1)$, where $(L_{1}, U_{1})$ is the PD interval for one randomly selected submatrix. Note that $(L,U)=(L_{1},U_{1})$ for $\rho_{(\ell)}$, $\ell=1,...,L$, since there is only one variation of the largest possible submatrix for $\rho_{(\ell)}$ (explanation in Section \ref{Supplement - Building the largest}). We average the M-H acceptance rates across the 500 datasets.
	\\ \\
	The M-H acceptance rates of correlations drawn from different candidate distributions are given in Tables \ref{table:sim4}, \ref{table:sim5}, and \ref{table:sim6} in the \hyperref[Supplement A]{Supporting Information}. For the correlations drawn from a reparameterized-beta distribution, acceptance rates are between 25-26\% after adjusting the tuning parameters. For correlations drawn from $\text{Unif}(L,U)$, the acceptance rates are between 5.4-8.0\% for the $\eta$'s, between 4.5-7.6\% for the $\rho$'s, and less than or equal to 2.3\% for $\gamma$. For correlations drawn from $\text{Unif}(L_{1},U_{1})$, the acceptance rates are between 5.2-7.9\% for the $\eta$'s and 2.0\% for $\gamma$. For the correlations drawn from $\text{Unif}(-1,1)$, the acceptance rates are between 4.0-6.4\% for the $\eta$'s, between 3.6-5.8\% for the $\rho$'s, and less than or equal to 1.9\% for $\gamma$. These results are as expected.
	\\ \\
	Next, we consider the percentage of times a correlation candidate resulted in a PD correlation matrix. We average the positive definiteness rates across the 500 datasets. The positive definiteness rates of correlations drawn from different candidate distributions are given in Tables \ref{table:sim7}, \ref{table:sim8}, and \ref{table:sim9} in the \hyperref[Supplement A]{Supporting Information}. In terms of positive definiteness rates, correlations drawn from $\text{R-Beta}(L,U)$ are the highest and $\text{Unif}(-1,1)$ the lowest as expected. When $\eta$'s are drawn from a reparameterized-beta distribution or when $\rho$'s are drawn from $\text{Unif}(L,U)$, their positive definiteness rates are consistent across the ambulatory disease stages.
	\\ \\
	It is apparent from Tables \ref{table:sim7}, \ref{table:sim8}, and \ref{table:sim9} that tuning the reparameterized-beta distribution to achieve 25\% acceptance rates in the M-H has a noticeable impact on the positive definiteness rates of the correlations. This is evident by the smaller differences in the positive definiteness rates between $\text{R-Beta}(L,U)$ and $\text{R-Beta}(-1,1)$, compared to the corresponding differences between $\text{Unif}(L,U)$ and $\text{Unif}(-1,1)$. Using one randomly selected submatrix to compute the PD interval $(L_{1}, U_{1})$ is computationally efficient and yields competitive but smaller positive definiteness rates than the intersected PD interval $(L,U)$. Unlike the uniform distribution, the reparameterized-beta distribution is centered on the previous iteration value of a correlation element, a value shaped by the prior to guarantee a PD correlation matrix. Therefore, tightening the support of the reparameterized-beta distribution typically increases autocorrelation between samples across iterations.
	\\ \\
	We note that a resulting structured correlation matrix from a correlation candidate is more likely to be PD if the correlation has a lower count of instances in the matrix. For example, in our structured correlation matrix $\bm{R}$, there are $2J$ instances of an $\eta_{\ell\ell'}$, $J(J-1)$ instances of a $\rho_{(\ell)}$, and $JL(J-1)(L-1)$ instances of $\gamma$. So if we were to draw correlations from $\text{Unif}(-1,1)$, i.e. without influence from PD intervals or tuning parameters, we expect $\gamma$ to have the lowest rate of producing a PD $\bm{R}$, and for $J > 3$, we expect $\eta$'s to have higher rates than $\rho$'s. This intuition holds in Tables \ref{table:sim7}, \ref{table:sim8}, and \ref{table:sim9} when correlations are drawn from a uniform distribution. In the case of the reparameterized-beta distribution, the positive definiteness rates of some $\rho$'s are slightly smaller than that of $\gamma$ due to tuning parameters.
	\\ \\
	Some further intuition concerning the positive definiteness rates of correlations drawn from candidate distributions supported on the PD interval is that the dimension of the largest possible submatrix for $\rho_{(\ell)}$ is 1 greater than that for $\eta_{\ell\ell'}$, but the submatrix for $\rho_{(\ell)}$ has no other variations, whereas the submatrix for $\eta_{\ell\ell'}$ has $J$ variations and thus $J$ PD intervals to intersect for the support of the $\eta_{\ell\ell'}$ candidate. And while the dimension of the largest possible submatrix for $\gamma$ is only 3, there are $2 {L \choose 2}$ PD intervals to intersect for the support of the $\gamma$ candidate.
	
	\section{Analysis of DMD biomarkers} \label{Data analysis}
	We use our model to make inference on optimal combination of biomarkers in the DMD data with respect to SRM. Recall from Section \ref{Data and motivation}, we have four muscles of interest, and we measure their annualized FF change between visits. We focus on two lower extremity muscles, SOL and VL, and two upper extremity muscles, BB and DEL. The data is stratified into three ambulatory disease stages.
	
	\subsection{Computations} \label{Data analysis - Computations}
	We run 4 MCMC chains, each with 75000 iterations and 1000 burn-in. The Metropolis-Hastings (M-H) acceptances rates of correlations drawn from different candidate distributions are given in Tables \ref{table:DA1}, \ref{table:DA2}, and \ref{table:DA3} in the \hyperref[Supplement A]{Supporting Information}. Across ambulatory disease stages, there is more variation among the acceptance rates of correlations here than those from the simulations (Section \ref{Acceptance and positive definiteness rates of correlations}). This is most likely because the simulated data is balanced without missingness and has a relatively small number of time points $J=4$. The DMD data is unbalanced with missingness and has varying distributions of the number of subject measurement times for different ambulatory disease stages (Tables \ref{table:DM1}, \ref{table:DM2}, and \ref{table:DM3} in the \hyperref[Supplement A]{Supporting Information}).
	\\ \\
	Next, the positive definiteness rates of correlations drawn from different candidate distributions are given in Tables \ref{table:DA4}, \ref{table:DA5}, and \ref{table:DA6} in the \hyperref[Supplement A]{Supporting Information}. As explained in in Section \ref{Acceptance and positive definiteness rates of correlations}, when drawing from $\text{Unif}(-1,1)$, we expect $\gamma$ to have the lowest rate of producing a PD $\bm{R}$, and for $J > 3$, we expect $\eta$'s to have higher rates than $\rho$'s. These expectations are largely supported by the results in Tables \ref{table:DA4}, \ref{table:DA5}, and \ref{table:DA6}, including for correlations drawn from $\text{R-Beta}(L,U)$ or $\text{Unif}(L,U)$. The only exceptions occur in the late ambulatory data, where not all positive definiteness rates of $\eta$'s exceed those of $\rho$'s under $\text{Unif}(-1,1)$  (and $\text{R-Beta}(L,U)$ to a much lesser extent). Differences in positive definiteness rates between $\eta$'s and $\rho$'s may be less pronounced in the late ambulatory group because 54.1\% of subjects had only one total measurement time, compared to 25.7\% for early ambulatory and 39.2\% for non-ambulatory subjects (Table \ref{table:DM3}).
	\\ \\
	The positive definiteness rates are lower here than those from the simulations (Section \ref{Acceptance and positive definiteness rates of correlations}). While the simulated data has a relatively small number of time points $J=4$, the maximum of subject measurement times for the DMD data is $J=8$, $6$, $7$ (early, late, non, respectively). The DMD data is also incomplete as opposed to the simulated data (Tables \ref{table:DM1}, \ref{table:DM2}, and \ref{table:DM3}).
	
	\subsection{Posteriors of model parameters} \label{Posteriors of model parameters}
	Posterior distributions of model parameters $\tilde{\bm{\mu}}$, $\bm{s}$, and $\bm{r}$ are summarized using 95\% credible intervals and posterior medians. These are given in Tables \ref{table:A2}, \ref{table:A3}, \ref{table:A4}, \ref{table:A5}, \ref{table:A6}, \ref{table:A7}, \ref{table:DA7}, \ref{table:DA8}, and \ref{table:DA9} in the \hyperref[Supplement A]{Supporting Information}.
	\\ \\
	The means of lower extremities increase from early to late ambulatory and then decrease from late to non-ambulatory, whereas the means of upper extremities increase from early to non-ambulatory. Note however that DEL shows little to no change from late to non-ambulatory. The means of upper extremities become greater than the means of lower extremities by the non-ambulatory disease stage. These patterns suggest that DMD affects the lower extremities of individuals more (larger average annualized changes of FF) for early and late ambulatory disease stages. By the time they cannot walk, DMD affects the upper extremities more, particularly BB.
	\\ \\
	The standard deviations increase from early to non-ambulatory, i.e. there is more variability in the annualized changes of FF in the muscles for later ambulatory disease stages. From early to non-ambulatory, VL has the highest standard deviation, and SOL has the lowest standard deviation.
	\\ \\
	Most of the correlations are positive, and they decrease as the individuals' ability to walk deteriorates. This is particularly apparent when comparing just the early and non-ambulatory disease stages. Generally, there are less relations between two muscles or two measurement times at later ambulatory disease stages in terms of annualized changes of FF in the muscles. Note that $\rho_{(1)}$, $\rho_{(2)}$, $\rho_{(3)}$, $\gamma$ have near 0 correlations at later ambulatory disease stages, but $\rho_{(4)}$ keeps a high correlation relative to other temporal correlations at later ambulatory disease stages. Between different measurement times, the annualized changes of FF become more similar for a lower extremity by the late ambulatory disease stage and for BB by the non-ambulatory disease stage. A possible explanation is that by the late ambulatory disease stage, DMD has progressed in the lower extremities to the point that there is a similar amount of FF replacement between different measurement times, and likewise for BB by the non-ambulatory disease stage. This is consistent with DMD targeting the lower extremities first.
	
	\subsection{Posteriors of weights and SRM} \label{Data analysis - Posteriors of weights and SRM}
	Table \ref{table:DAweights&SRMs} provides the point estimate of the weights, which is the set of weights that optimize the SRM, by plugging in the posterior medians of the model parameters (Table \ref{table:DA10} in the \hyperref[Supplement A]{Supporting Information}). Posterior density is given in Figure \ref{w_dens.png} in the \hyperref[Supplement A]{Supporting Information}. Credible intervals are given in Tables \ref{table:DA11}, \ref{table:DA12}, and \ref{table:DA13} in the \hyperref[Supplement A]{Supporting Information}. The estimates of the individual muscle SRMs, the SRM using optimal weights ($\text{SRM}_{\text{opt}}$), and the SRM using equal weights $\bm{w}_{\text{equal}}=(0.25,0.25,0.25,0.25)'$ ($\text{SRM}_{\text{equal}}$) are also given in Table \ref{table:DAweights&SRMs}. $\text{SRM}_{\text{equal}}$ uses the same model parameters as $\text{SRM}_{\text{opt}}$. Note that $\text{SRM}_{\text{opt}}$ and $\text{SRM}_{\text{equal}}$ account for correlations between muscles unlike the individual muscle SRMs.
	\\ \\
	SOL weight increases from early to late ambulatory and then decreases from late to non-ambulatory. VL weight decreases from early to non-ambulatory. BB weight very slightly decreases from early to late ambulatory and then greatly increases from late to non-ambulatory. DEL weight is zero or close to zero across the ambulatory disease stages which may be explained by its relatively high correlation with other muscles, particularly SOL and BB (Table \ref{table:DA10}). Among the four muscles, the lower extremities and BB are the most responsive muscles to FF replacement across all ambulatory disease stages. As the individuals lose their ability to walk, BB becomes increasingly more responsive and eventually contributes the most weight by the non-ambulatory disease stage. At early ambulatory, VL contributes more weight than SOL, but this eventually shifts by late ambulatory and persists to non-ambulatory.
	\\ \\ 
	The trends for individual muscle SRMs across the ambulatory disease stages are similar to the observed trends from before for the point estimate of the weights. As expected, $\text{SRM}_{\text{opt}}$ is greater than any individual muscle SRM and $\text{SRM}_{\text{equal}}$. We note that VL SRM is greater than $\text{SRM}_{\text{equal}}$ at early ambulatory, and BB SRM is greater than $\text{SRM}_{\text{equal}}$ at non-ambulatory.
	\\ \\
	Table \ref{table:SOL&VLmodel} provides corresponding estimates as in Table \ref{table:DAweights&SRMs} if the model only considers SOL and VL. Individual muscle SRMs for SOL and VL are the same as in Table \ref{table:DAweights&SRMs}. We find similar trends for the point estimate of the weights with the exception that SOL shows only an increasing trend from early to non-ambulatory since no upper extremities are present. The SRM from optimal weights are smaller across the ambulatory disease stages when the model only considers lower extremities, particularly at non-ambulatory.
	\\ \\
	We visualize the joint posterior distribution of the weights which live on the simplex by plotting each set of weights as a point on a tetrahedron and then using a three-dimensional version of the boxplot called \texttt{gemPlot} (\texttt{R} package) \cite{Kruppa}. See Section \ref{Supplement B -- Gemplots} of the \hyperref[Supplement A]{Supporting Information} for further details and figures.
	
	\section{Discussion} \label{Discussion}
	Our objective was to estimate an optimal combination of biomarkers in order to assess disease progression of individuals with DMD. To do this we modelled the data using a multivariate normal distribution with a structured correlation matrix to account for the high rate of missingness. For Bayesian inference, we addressed the positive definiteness constraint of a structured correlation matrix by proposing a generalization of the interval approach in \cite{Barnard}. In particular, we adapted the Barnard approach to the set of the largest possible submatrices of a structured correlation matrix, where a target correlation parameter is a unique element inside each submatrix. This procedure computes a `tight' interval for the support of a correlation parameter. We provided detailed rationale and specific algorithms on how to build the largest possible submatrices in the context of our parameterization. We note that our approach can be used for any correlation structure without necessitating an ordering of variables including partial time series structures. For example, to replace the exchangeable structure with an autoregressive order one, we would replace $\gamma$ with $\gamma^{|\mbox{lag}|}$.
	\\ \\
	Our approach for the correlation structure in the multivariate normal model could also be used for a $t$ or skew-normal model. This was demonstrated in the simulations where we fit the multivariate normal model for datasets generated by $t_{10}$, $t_{3}$, and skew-normal distributions. MSEs and coverages of credible intervals were as expected and magnitudes of biases were small for the weights and standardized response mean of the construct. The simulations also showed that the positive definiteness and acceptance rates of the correlation parameters in MCMC were improved when using our approach. We compared different candidate distributions with tightened supports for the correlations parameters against $\text{Unif}(-1,1)$.
	\\ \\
	For each posterior sample of the model parameters, we computed optimal weights for the construct to evaluate DMD progression across different ambulatory disease stages. We also demonstrated how to visualize the joint posterior of the weights on the simplex. We found that at the early and late ambulatory disease stages, the lower extremities were the most responsive muscles. After the individuals lost their ability to walk, biceps brachii became the most responsive muscle. The posterior means and standardized response means of individual muscles also support our findings. As the disease progressed, variability of individual muscles increased, whereas correlations between muscles or measurement times generally decreased. We note that the deltoid muscle had a weight close to zero and has a relatively high correlation with other muscles across all disease stages. Fat fraction of the deltoid is difficult to measure using magnetic resonance spectroscopy due to heterogeneous distribution of fat across the muscle, making it difficult to capture a representative sample in a rectangular voxel.
	\\ \\
	We note that the constructs can vary with other disease characteristics besides ambulatory status, including other functional milestones or continuous outcomes, e.g. 6MWD (6 minute walking distance); we are currently working on this extension. We will also explore alternative objective functions besides the standardized response mean to optimize the weights. In addition, our approach to compute `almost' PD intervals can be used for concentration matrices.  Finally, we could consider a `composite' proposal, that first samples from the reparameterized-beta and then samples from the uniform based on our intervals to increase the acceptance rate and allow for a more efficient exploration of the posterior.
	

	\section*{Acknowledgments}
 All co-authors were partially supported by either NIH R01 AR056973
and/or Wellstone grant P50 (P50AR052646).  Daniels was also partially supported by NIH R01 HL158963.  MRI data was collected in the McKnight Brain Institute at the National High Magnetic Field Laboratory’s Advanced Magnetic Resonance Imaging and Spectroscopy (AMRIS) Facility, which is supported by National Science Foundation Cooperative Agreement No. DMR-1644779 and DMR-1157490 and the State of Florida, and in OHSU’s Advanced Imaging Research Center, supported by NIH S10OD021701 for the 3T Siemens Prisma MRI instrument.
Two instrument grants at OHSU also supported the data collection:  NIH S10-OD018224 and S10-OD016356.

	\section*{Conflicts of Interest}
	The authors declare no conflicts of interest.
	
	\section*{Data Availability Statement}
	Research data are not shared.
	
	\section*{Supporting Information}
	Section \ref{Supplement A} of the \hyperref[Supplement A]{Supporting Information} provides the sampling procedure, prior specifications, and other configurations of our MCMC algorithm. Section \ref{Supplement B} of the \hyperref[Supplement A]{Supporting Information} presents supplementary tables and figures for Sections \ref{Data and motivation}, \ref{Simulations}, and \ref{Data analysis}.
	\\ \\
	Our code is on GitHub at \url{https://github.com/michaelkkim/dmd_project1}. We provided an illustrative example to help others compute approximate positive definite bounds for their own structured correlation matrix.

	
	\bibliographystyle{wileyNJD-AMA.bst} 
	\bibliography{bib_dmd1.bib} 

\begin{thebibliography}{10}
\providecommand \doibase [0]{http://dx.doi.org/}%

\bibitem{Barnard}
Barnard J, McCulloch R, Meng XL. Modeling covariance matrices in terms of standard deviations and correlations, with application to shrinkage. {\it Statistica Sinica.} 2000\string:1281--1311.

\bibitem{Lindstrom}
Lindstrom MJ, Bates DM. Newton—Raphson and EM algorithms for linear mixed-effects models for repeated-measures data. {\it Journal of the American Statistical Association.} 1988\string;83(404)\string:1014--1022.

\bibitem{Leonard}
Leonard T, Hsu JS. Bayesian inference for a covariance matrix. {\it The Annals of Statistics.} 1992\string;20(4)\string:1669--1696.

\bibitem{Chiu}
Chiu TY, Leonard T, Tsui KW. The matrix-logarithmic covariance model. {\it Journal of the American Statistical Association.} 1996\string:198--210.

\bibitem{Yang}
Yang R, Berger JO. Estimation of a covariance matrix using the reference prior. {\it The Annals of Statistics.} 1994\string:1195--1211.

\bibitem{DanielsKass1999}
Daniels MJ, Kass RE. Non-conjugate Bayesian estimation of covariance matrices and its use in hierarchical models. {\it Journal of the American Statistical Association.} 1999\string;94(448)\string:1254--1263.

\bibitem{Pourahmadi1999}
Pourahmadi M. Joint mean-covariance models with applications to longitudinal data: Unconstrained parameterisation. {\it Biometrika.} 1999\string;86(3)\string:677--690.

\bibitem{Pan2003}
Pan J, Mackenzie G. On modelling mean-covariance structures in longitudinal studies. {\it Biometrika.} 2003\string;90(1)\string:239--244.

\bibitem{Pan2006}
Pan J, MacKenzie G. Regression models for covariance structures in longitudinal studies. {\it Statistical Modelling.} 2006\string;6(1)\string:43--57.

\bibitem{Leng2010}
Leng C, Zhang W, Pan J. Semiparametric mean--covariance regression analysis for longitudinal data. {\it Journal of the American Statistical Association.} 2010\string;105(489)\string:181--193.

\bibitem{ZhangLeng2012}
Zhang W, Leng C. A moving average Cholesky factor model in covariance modelling for longitudinal data. {\it Biometrika.} 2012\string;99(1)\string:141--150.

\bibitem{Pinheiro}
Pinheiro JC, Bates DM. Unconstrained parametrizations for variance-covariance matrices. {\it Statistics and computing.} 1996\string;6\string:289--296.

\bibitem{Zhang}
Zhang W, Leng C, Tang CY. A joint modelling approach for longitudinal studies. {\it Journal of the Royal Statistical Society Series B: Statistical Methodology.} 2015\string;77(1)\string:219--238.

\bibitem{Tsay}
Tsay RS, Pourahmadi M. Modelling structured correlation matrices. {\it Biometrika.} 2017\string;104(1)\string:237--242.

\bibitem{Ghosh}
Ghosh RP, Mallick B, Pourahmadi M. Bayesian estimation of correlation matrices of longitudinal data. {\it Bayesian Analysis.} 2021\string;16(3)\string:1039--1058.

\bibitem{Daniels2009}
Daniels MJ, Pourahmadi M. Modeling covariance matrices via partial autocorrelations. {\it Journal of Multivariate Analysis.} 2009\string;100(10)\string:2352--2363.

\bibitem{Wang2013}
Wang Y, Daniels MJ. Bayesian modeling of the dependence in longitudinal data via partial autocorrelations and marginal variances. {\it Journal of Multivariate Analysis.} 2013\string;116\string:130--140.

\bibitem{Archakov}
Archakov I, Hansen PR. A new parametrization of correlation matrices. {\it Econometrica.} 2021\string;89(4)\string:1699--1715.

\bibitem{Hu}
Hu J, Chen Y, Leng C, Tang CY. Regression Analysis of Correlations for Correlated Data. {\it arXiv preprint arXiv:2109.05861.} 2021.

\bibitem{Anderson}
Anderson TW. Asymptotically efficient estimation of covariance matrices with linear structure. {\it The Annals of Statistics.} 1973\string;1(1)\string:135--141.

\bibitem{Wong}
Wong F, Carter CK, Kohn R. Efficient estimation of covariance selection models. {\it Biometrika.} 2003\string;90(4)\string:809--830.

\bibitem{Liechty}
Liechty JC, Liechty MW, M{\"u}ller P. Bayesian correlation estimation. {\it Biometrika.} 2004\string;91(1)\string:1--14.

\bibitem{Dempster}
Dempster AP. Covariance selection. {\it Biometrics.} 1972\string:157--175.

\bibitem{Pitt}
Pitt M, Chan D, Kohn R. Efficient Bayesian inference for Gaussian copula regression models. {\it Biometrika.} 2006\string;93(3)\string:537--554.

\bibitem{Carter}
Carter CK, Wong F, Kohn R. Constructing priors based on model size for nondecomposable Gaussian graphical models: A simulation based approach. {\it Journal of Multivariate Analysis.} 2011\string;102(5)\string:871--883.

\bibitem{Zhang2022}
Zhang S, Kuha J, Steele F. Modelling Correlation Matrices in Multivariate Dyadic Data: Latent Variable Models for Intergenerational Exchanges of Family Support. {\it arXiv preprint arXiv:2210.14751.} 2022.

\bibitem{Forbes}
Forbes SC, Arora H, Willcocks RJ, et al. Upper and lower extremities in Duchenne muscular dystrophy evaluated with quantitative MRI and proton MR spectroscopy in a multicenter cohort. {\it Radiology.} 2020\string;295(3)\string:616--625.

\bibitem{Verhaart}
Verhaart IE, Aartsma-Rus A. Therapeutic developments for Duchenne muscular dystrophy. {\it Nature Reviews Neurology.} 2019\string;15(7)\string:373--386.

\bibitem{Gelman}
Gelman A, Gilks WR, Roberts GO. Weak convergence and optimal scaling of random walk Metropolis algorithms. {\it The annals of applied probability.} 1997\string;7(1)\string:110--120.

\bibitem{Kotz}
Kotz S, Nadarajah S. {\it Multivariate t-distributions and their applications}.
\newblock Cambridge University Press, 2004.

\bibitem{Azzalini1999}
Azzalini A, Capitanio A. Statistical applications of the multivariate skew normal distribution. {\it Journal of the Royal Statistical Society: Series B (Statistical Methodology).} 1999\string;61(3)\string:579--602.

\bibitem{Kruppa}
Kruppa J, Jung K. Automated multigroup outlier identification in molecular high-throughput data using bagplots and gemplots. {\it BMC bioinformatics.} 2017\string;18(1)\string:1--10.

\bibitem{Taddy}
Taddy MA. {\it Bayesian nonparametric analysis of conditional distributions and inference for Poisson point processes}.
\newblock University of California, Santa Cruz, 2008.

\bibitem{Harville}
Harville DA. {\it Matrix Algebra From a Statistician's Perspective}.
\newblock Springer Science \& Business Media, 2008.

\bibitem{Deaux}
Deaux R. {\it Introduction to the geometry of complex numbers}.
\newblock Courier Corporation, 2013.

\bibitem{Rousseauw}
Rousseeuw PJ, Ruts I, Tukey JW. The bagplot: a bivariate boxplot. {\it The American Statistician.} 1999\string;53(4)\string:382--387.

\bibitem{Gower}
Gower JC, Lubbe SG, Le~Roux NJ. {\it Understanding biplots}.
\newblock John Wiley \& Sons, 2011.

\end{thebibliography}
	
	\newpage
	\begin{table}[H]
    \centering
        \begin{tabular}{l|ccccc}
        Coverage & $w_{1}$ & $w_{2}$ & $w_{3}$ & $w_{4}$ & SRM \\
        \hline
        Normal -- no miss -- early ambulatory & 95.2\% & 96.2\% & 95.6\% & 100\% & 94.8\% \\
        Normal -- no miss -- late ambulatory & 95.6\% & 95.4\% & 95.4\% & 95.0\% & 94.8\% \\
        Normal -- no miss -- non-ambulatory & 95.6\% & 96.8\% & 96.0\% & 99.6\% & 95.0\% \\
        \hline
        $t_{10}$ -- no miss -- early ambulatory & 91.6\% & 93.4\% & 94.2\% & 100\% & 92.6\% \\
        $t_{10}$ -- no miss -- late ambulatory & 93.8\% & 93.4\% & 93.0\% & 95.2\% & 90.6\% \\
        $t_{10}$ -- no miss -- non-ambulatory & 94.0\% & 94.2\% & 93.8\% & 99.4\% & 91.6\% \\
        \hline
        $t_{3}$ -- no miss -- early ambulatory & 77.2\% & 76.2\% & 78.2\% & 99.2\% & 56.6\% \\
        $t_{3}$ -- no miss -- late ambulatory & 74.6\% & 72.8\% & 78.6\% & 78.6\% & 55.0\% \\
        $t_{3}$ -- no miss -- non-ambulatory & 75.8\% & 76.8\% & 75.0\% & 96.2\% & 50.8\% \\
        \hline
        Skew-normal -- no miss -- early ambulatory & 88.6\% & 73.8\% & 93.4\% & 99.8\% & 19.0\% \\
        Skew-normal -- no miss -- late ambulatory & 89.2\% & 94.8\% & 95.0\% & 90.6\% & 20.0\% \\
        Skew-normal -- no miss -- non-ambulatory & 96.0\% & 93.8\% & 82.2\% & 97.2\% & 39.2\% \\
        \hline
        Normal -- yes miss -- early ambulatory & 95.8\% & 95.8\% & 95.2\% & 100\% & 95.2\% \\
        Normal -- yes miss -- late ambulatory & 94.0\% & 94.8\% & 93.8\% & 94.6\% & 94.2\% \\
        Normal -- yes miss -- non-ambulatory & 95.2\% & 96.4\% & 94.8\% & 99.0\% & 94.2\% \\
        \end{tabular}
        \caption{Coverages of 95\% credible intervals of weights and SRM}
		\label{table:sim1}
    \end{table}
    
    \begin{table}[H]
    \centering
        \begin{tabular}{l|ccccc}
        Bias & $w_{1}$ & $w_{2}$ & $w_{3}$ & $w_{4}$ & SRM \\
        \hline
        Normal -- no miss -- early ambulatory & 0.00 & 0.00 & -0.00 & 0.00 & 0.01 \\
        Normal -- no miss -- late ambulatory & -0.00 & -0.00 & -0.00 & 0.01 & 0.00 \\
        Normal -- no miss -- non-ambulatory & -0.00 & -0.00 & -0.00 & 0.01 & -0.00 \\
        \hline
        $t_{10}$ -- no miss -- early ambulatory & 0.01 & -0.01 & -0.00 & 0.00 & 0.01 \\
        $t_{10}$ -- no miss -- late ambulatory & 0.00 & -0.00 & -0.01 & 0.01 & 0.00 \\
        $t_{10}$ -- no miss -- non-ambulatory & -0.00 & 0.00 & -0.01 & 0.01 & -0.01 \\
        \hline
        $t_{3}$ -- no miss -- early ambulatory & -0.01 & 0.01 & -0.00 & 0.01 & 0.08 \\
        $t_{3}$ -- no miss -- late ambulatory & -0.00 & -0.00 & -0.01 & 0.02 & 0.08 \\
        $t_{3}$ -- no miss -- non-ambulatory & -0.02 & 0.01 & -0.01 & 0.02 & 0.09 \\
        \hline
        Skew-normal -- no miss -- early ambulatory & 0.06 & -0.08  & 0.02 & 0.00 & 0.25 \\
        Skew-normal -- no miss -- late ambulatory & -0.03 & -0.01 & 0.00 & 0.04 & 0.21 \\
        Skew-normal -- no miss -- non-ambulatory & -0.00 & 0.02 & -0.04 & 0.02 & 0.15 \\
        \hline
        Normal -- yes miss -- early ambulatory & -0.00 & 0.00 & -0.01 & 0.01 & 0.01 \\
        Normal -- yes miss -- late ambulatory & -0.01 & -0.00 & -0.01 & 0.03 & 0.01 \\
        Normal -- yes miss -- non-ambulatory & -0.01 & 0.00 & -0.02 & 0.03 & 0.00 \\
        \end{tabular}
        \caption{Biases of weights and SRM}
		\label{table:sim2}
    \end{table}
    
    \begin{table}[H]
	\centering
		\begin{tabular}{l|ccc}
			& Early ambulatory & Late ambulatory & Non-ambulatory \\
			\hline
			$w_{1}$[SOL] & 0.290 & 0.459 & 0.358 \\
			$w_{2}$[VL] & 0.465 & 0.218 & 0.051 \\
			$w_{3}$[BB] & 0.245 & 0.235 & 0.591 \\
			$w_{4}$[DEL] & 0 & 0.088 & 0 \\
			\hline
			$\text{SRM}_{\text{SOL}}$ & 0.706 & 0.957 & 0.706 \\
			$\text{SRM}_{\text{VL}}$ & 0.913 & 0.804 & 0.335 \\
			$\text{SRM}_{\text{BB}}$ & 0.563 & 0.687 & 1.010 \\
			$\text{SRM}_{\text{DEL}}$ & 0.368 & 0.770 & 0.567 \\
			\hline
			$\text{SRM}_{\text{opt}}$ & 0.985 & 1.183 & 1.144 \\
			\hline
			$\text{SRM}_{\text{equal}}$ & 0.901 & 1.137 & 0.940 \\
		\end{tabular}
		\caption{Point estimate of weights, individual muscle SRMs, SRM from optimal weights, SRM from equal weights}
		\label{table:DAweights&SRMs}
	\end{table}
	
	\begin{table}[H]
	\centering
		\begin{tabular}{l|ccc}
			& Early ambulatory & Late ambulatory & Non-ambulatory \\
			\hline
			$w_{1}$[SOL] & 0.392 & 0.665 & 0.936 \\
			$w_{2}$[VL] & 0.608 & 0.335 & 0.064 \\
			\hline
			$\text{SRM}_{\text{opt}}$ & 0.935 & 1.098 & 0.708 \\
			\hline
			$\text{SRM}_{\text{equal}}$ & 0.932 & 1.065 & 0.579 \\
		\end{tabular}
		\caption{Point estimate of weights, SRM from optimal weights, SRM from equal weights (for model with only SOL and VL)}
		\label{table:SOL&VLmodel}
	\end{table}
	
	\newpage
	\thispagestyle{empty}
	\ \newpage
	
	\begin{center}
  		{\Huge Supporting Information}
	\end{center}

	\appendix  
	\renewcommand{\thesection}{S\Alph{section}}
	\renewcommand{\thepage}{S\arabic{page}}
	\setcounter{page}{1}
	\renewcommand{\thealgocf}{S\Alph{section}\arabic{algocf}}
	\setcounter{algocf}{0}
	\renewcommand{\thetable}{S\Alph{section}\arabic{table}}
	\setcounter{table}{0}
	\renewcommand{\thefigure}{S\Alph{section}\arabic{figure}}
	\setcounter{figure}{0}

	\section{Supporting Information A} \label{Supplement A}
	\subsection{Details of MCMC algorithm} \label{Supplement - Details of MCMC algorithm}
	\begin{itemize}[nosep]
		\item MCMC has 4 chains for non-simulated data, each with $T=75000$ iterations and $1000$ burn-in. For the 500 simulated datasets, we use $T=50000$ iterations and $1000$ burn-in
		\item Other than when adapting the Barnard approach to compute the positive definite (PD) intervals to sample $\bm{r}$, everything else in the algorithm is invariant to the ordering of $\bm{y}_{i}$ and uses the following ordering:
		$$\underbrace{\bm{y}_{i}}_{p \times 1 = JL \times 1}=(y_{i11}, ..., y_{i1L}, \ \ \ ... \ \ \ , y_{iJ1}, ..., y_{iJL})'.$$
		\item For initializing the algorithm for posterior computation, set the correlations to $0$, and use the sample means and sample standard deviations of the muscles:
		\begin{itemize}[nosep]
			\item $\mu_{\ell}^{(0)}\stackrel{\text{set}}{=}\frac{1}{p_{\ell,{\text{obs}}}}\sum_{i=1}^{N}\sum_{j=1}^{J}y_{ij\ell}\mathds{1}\{y_{ij\ell} \text{ observed}\}=\bar{y}_{\ell,{\text{obs}}}, \ \ \ \ell=1,...,L,$ \\
			where $p_{\ell,{\text{obs}}}=\sum_{i=1}^{N}\sum_{j=1}^{J}\mathds{1}\{y_{ij\ell} \text{ observed}\}$.
			\item $s_{\ell}^{2(0)}\stackrel{\text{set}}{=}\frac{1}{p_{\ell,{\text{obs}}}-1}\sum_{i=1}^{N}\sum_{j=1}^{J}(y_{ij\ell}\mathds{1}\{y_{ij\ell} \text{ observed}\}-\bar{y}_{\ell,{\text{obs}}})^{2}=\hat{s}_{\ell,{\text{obs}}}^{2}, \ \ \ \ell=1,...,L$.
			\item $r_{k}^{(0)}\stackrel{\text{set}}{=}0, \ \ \ k=1,...,q, \ \ \ $ i.e. $\bm{R}^{(0)}=\bm{I}$.
			\item $\bm{\Sigma}^{(0)}=\bm{S}^{(0)}\bm{R}^{(0)}\bm{S}^{(0)}$.
		\end{itemize}
		\item for $(t$ in $1:T)$ \{
		\begin{itemize}[nosep]
			\item $\tilde{\bm{\mu}}^{(t)}|\bm{y}_{\text{obs}},\bm{\Sigma}^{(t-1)}$
			\\
			$\sim\text{MVN}\Big(\Big[\sum_{i=1}^{N}\bm{X}_{i,\text{obs}}'\bm{\Sigma}^{-1(t-1)}_{\text{obs}_{i},\text{obs}_{i}}\bm{X}_{i,\text{obs}}+\bm{\Sigma}_{0}^{-1}\Big]^{-1}\Big[\sum_{i=1}^{N}\bm{X}_{i,\text{obs}}'\bm{\Sigma}^{-1(t-1)}_{\text{obs}_{i},\text{obs}_{i}}\bm{y}_{i,\text{obs}}+\bm{\Sigma}_{0}^{-1}\bm{\mu}_{0}\Big],$
			\\
			$\Big[\sum_{i=1}^{N}\bm{X}_{i,\text{obs}}'\bm{\Sigma}^{-1(t-1)}_{\text{obs}_{i},\text{obs}_{i}}\bm{X}_{i,\text{obs}}+\bm{\Sigma}_{0}^{-1}\Big]^{-1} \Big)$ (Section \ref{Conditional posterior of means}),  where
			$$\tilde{\bm{\mu}}^{(t)}=\begin{bsmallmatrix}
				\mu_{1}^{(t)} \\
				\vdots \\
				\mu_{L}^{(t)} \\
			\end{bsmallmatrix}, \ \ \ \bm{X}_{i,\text{obs}}=\begin{bsmallmatrix}
				\mathds{1}\{\ell=1\} & ... & \mathds{1}\{\ell=L\} \\
				\vdots \\
				\mathds{1}\{\ell=1\} & ... & \mathds{1}\{\ell=L\}
			\end{bsmallmatrix}_{p_{i,\text{obs} } \times L}, \ \ \  \bm{\Sigma}_{\text{obs}_{i},\text{obs}_{i}} \text { is } \ p_{i,\text{obs}} \times p_{i,\text{obs}},$$
			$$p_{i,\text{obs}}=\sum_{j=1}^{J}\sum_{\ell=1}^{L} \mathds{1}\{y_{ij\ell} \text{ observed}\},$$
			$$\bm{\mu}_{0}=(\bar{y}_{1,{\text{obs}}},...,\bar{y}_{L,{\text{obs}}})', \ \ \ \bm{\Sigma}_{0}=\text{diag}\bigg(\bigg(\frac{\text{range}(\bm{y}_{1,{\text{obs}}})}{4}\bigg)^{2},...,\bigg(\frac{\text{range}(\bm{y}_{L,{\text{obs}}})}{4}\bigg)^{2}\bigg),$$
			$$\text{range}(\bm{y}_{\ell,{\text{obs}}})=\text{max}\{y_{ij\ell}  : y_{ij\ell} \text{ observed}\}-\text{min}\{y_{ij\ell} : y_{ij\ell} \text{ observed}\}.$$
			\item Update $\bm{\mu}^{(t)}=(\mu_{1}^{(t)},...,\mu_{L}^{(t)}, \ \ \ ... \ \ \ , \mu_{1}^{(t)},...,\mu_{L}^{(t)})'$.
			\item for $(\ell$ in $1:L)$ \{
			\begin{itemize}[nosep]
				\item $s_{\ell}^{2(t)} \sim\text{Inv-Gamma}\Big(2.1+\frac{NJ}{2},$
				\\
				$3.1\hat{s}_{\ell,{\text{obs}}}^{2}+ \frac{1}{2}\sum_{i=1}^{N}\sum_{j=1}^{J}(y_{ij\ell}\mathds{1}\{y_{ij\ell} \text{ observed}\} -\mu_{\ell}^{(t)})^{2}\Big)$ (Section \ref{Candidate distribution of lth variance}).
				\item  Compute $\alpha_{s} = \min\bigg\{1, \frac{l(\bm{y}_{\text{obs}}|s_{\ell}^{(t)}, \bm{s}_{(-\ell)}^{(t-1)}, \bm{\mu}^{(t)}, \bm{r}^{(t-1)}) p(s_{\ell}^{2(t)}) / f_{c}(s_{\ell}^{2(t)})}{l(\bm{y}_{\text{obs}}|s_{\ell}^{(t-1)}, \bm{s}_{(-\ell)}^{(t-1)}, \bm{\mu}^{(t)}, \bm{r}^{(t-1)}) p(s_{\ell}^{2(t-1)}) / f_{c}(s_{\ell}^{2(t-1)})}\bigg\}$, where
				\begin{itemize}[nosep]
					\item $l(\bm{y}_{\text{obs}}|s_{\ell}^{(t)},\bm{s}_{(-\ell)}^{(t-1)},\bm{\mu}^{(t)},\bm{r}^{(t-1)})=\text{density-MVN}(\bm{y}_{\text{obs}}; \bm{\mu}^{(t)},\bm{\Sigma}(s_{\ell}^{(t)},\bm{s}_{(-\ell)}^{(t-1)}, \bm{r}^{(t-1)}))$, where \\
					$\bm{\Sigma}(s_{\ell}^{(t)}, \bm{s}_{(-\ell)}^{(t-1)}, \bm{r}^{(t-1)})=\bm{S}(s_{\ell}^{(t)}, \bm{s}_{(-\ell)}^{(t-1)})\bm{R}^{(t-1)}\bm{S}(s_{\ell}^{(t)}, \bm{s}_{(-\ell)}^{(t-1)})$, where \\
					$\bm{R}^{(t-1)}$ is the structured correlation matrix with unique elements $\bm{r}^{(t-1)}$, and \\
					$\bm{S}(s_{\ell}^{(t)}, \bm{s}_{(-\ell)}^{(t-1)})$ is the structured diagonal standard deviation matrix with unique elements $s_{\ell}^{(t)}, \bm{s}_{(-\ell)}^{(t-1)}$, where $\bm{s}_{(-\ell)}^{(t-1)}=(s_{1}^{(t)}, ..., s_{\ell-1}^{(t)}, s_{\ell+1}^{(t-1)}, ..., s_{L}^{(t-1)})'$.
					\item $p(s_{\ell}^{2(t)}) = \text{density-Inv-Gamma}(s_{\ell}^{2(t)}; \nu_{\ell},(\nu_{\ell}+1)\hat{s}_{\ell,\text{obs}}^{2})$.
					\item $f_{c}(s_{\ell}^{2(t)})=\text{density-Inv-Gamma}\Big(s_{\ell}^{2(t)}; 2.1+\frac{NJ}{2},$
					\\
					$3.1\hat{s}_{\ell,\text{obs}}^{2}+\frac{1}{2}\sum_{i=1}^{N}\sum_{j=1}^{J}(y_{ij\ell}\mathds{1}\{y_{ij\ell} \text{ observed}\} -\mu_{\ell}^{(t)})^{2}\Big)$.
				\end{itemize}
				\item Accept $s_{\ell}^{(t)}$ with probability $\alpha_{s}$.
				\item Update $\bm{S}: s_{\ell}^{(t-1)} \rightarrow s_{\ell}^{(t)}$.
			\end{itemize}
			\}
			\item for $(k$ in $1:q)$ \{
			\begin{itemize}[nosep]
				\item Adapt the Barnard approach \cite{Barnard} (Section \ref{Supplement - Barnard approach details}) to the set of largest possible submatrices of $\bm{R}$ (Section \ref{Supplement - Building the largest}), where each of these submatrices contains the $k$th correlation $r_{k} \in (r_{1},...,r_{q})'$ only once, and the rest of the submatrix is filled with current values of correlation elements other than the $k$th corre lation $\bm{r}_{(-k)}^{(t-1)}$. After computing the PD interval for each submatrix, define the support of the candidate distribution of $r_{k}^{(t)}$ as the intersection of all the PD intervals. If there is only one largest possible submatrix, then use its PD interval as the support. The support is denoted as $(L_{k}^{(t)}, U_{k}^{(t)})$.
				\item $r_{k}^{(t)} \sim \text{Unif}(L_{k}^{(t)}, U_{k}^{(t)})$, or
				\\
				$r_{k}^{(t)} \sim \text{R-Beta}(\alpha_{k}^{(t)}, \beta_{k}^{(t)}, L_{k}^{(t)}, U_{k}^{(t)})$, where
				\begin{itemize}[nosep]
					\item $\text{Mode}[r_{k}^{(t)}]\stackrel{\text{set}}{=}r_{k}^{(t-1)} \ \ \ \Rightarrow$
					$$\alpha_{k}^{(t)}=\frac{(\kappa_{k}-1)L_{k}^{(t)}+(2-\kappa_{k})r_{k}^{(t-1)}-U_{k}^{(t)}}{L_{k}^{(t)}-U_{k}^{(t)}},$$
					$$\beta_{k}^{(t)}=\frac{(\kappa_{k}-1)U_{k}^{(t)}+(2-\kappa_{k})r_{k}^{(t-1)}-L_{k}^{(t)}}{U_{k}^{(t)}-L_{k}^{(t)}}=\kappa_{k}-\alpha_{k}^{(t)}.$$
					\item Adjust $\kappa_{k}$ to achieve about 25\% acceptance rate.
				\end{itemize}
				\item Compute $\alpha_{r} = \min\bigg\{1, \frac{l(\bm{y}_{\text{obs}}|r_{k}^{(t)}, \bm{r}_{(-k)}^{(t-1)}, \bm{\mu}^{(t)}, \bm{s}^{(t)}) p(r_{k}^{(t)}, \bm{r}_{(-k)}^{(t-1)}) / g_{c}(r_{k}^{(t)})}{l(\bm{y}_{\text{obs}}|r_{k}^{(t-1)}, \bm{r}_{(-k)}^{(t-1)}, \bm{\mu}^{(t)}, \bm{s}^{(t)}) p(r_{k}^{(t-1)}, \bm{r}_{(-k)}^{(t-1)}) / g_{c}(r_{k}^{(t-1)})}\bigg\}$, where
				\begin{itemize}[nosep]
					\item $l(\bm{y}_{\text{obs}}|r_{k}^{(t)}, \bm{r}_{(-k)}^{(t-1)}, \bm{\mu}^{(t)}, \bm{s}^{(t)})=\text{density-MVN}(\bm{y}_{\text{obs}}; \bm{\mu}^{(t)}, \bm{\Sigma}(r_{k}^{(t)}, \bm{r}_{(-k)}^{(t-1)}, \bm{s}^{(t)}))$, where \\
					$\bm{\Sigma}(r_{k}^{(t)}, \bm{r}_{(-k)}^{(t-1)}, \bm{s}^{(t)})=\bm{S}^{(t)}\bm{R}(r_{k}^{(t)}, \bm{r}_{(-k)}^{(t-1)})\bm{S}^{(t)}$, where \\
					$\bm{S}^{(t)}$ is the structured diagonal standard deviation matrix with unique elements $\bm{s}^{(t)}$, and \\
					$\bm{R}(r_{k}^{(t)}, \bm{r}_{(-k)}^{(t-1)})$ is the structured correlation matrix with unique elements $r_{k}^{(t)}, \bm{r}_{(-k)}^{(t-1)}$, where $\bm{r}_{(-k)}^{(t-1)}=(r_{1}^{(t)}, ..., r_{k-1}^{(t)}, r_{k+1}^{(t-1)}, ..., r_{q}^{(t-1)})'$.
					\item $p(r_{k}^{(t)}, \bm{r}_{(-k)}^{(t-1)}) = \mathds{1}\{\bm{R}(r_{k}^{(t)}, \bm{r}_{(-k)}^{(t-1)}) \text{ is P.D.})\}$ \cite{Liechty}. \\
					Note that by construction, $p(r_{k}^{(t-1)}, \bm{r}_{(-k)}^{(t-1)})=\mathds{1}\{\bm{R}(r_{k}^{(t-1)}, \bm{r}_{(-k)}^{(t-1)}) \text{ is P.D.}\}=1$.
					\item $g_{c}(r_{k}^{(t)}) = \text{density-Unif}(L_{k}^{(t)}, U_{k}^{(t)})$ or
					\\
					$g_{c}(r_{k}^{(t)} |r_{k}^{(t-1)})=\text{density-\text{R-Beta}}(r_{k}^{(t)}; \alpha_{k}^{(t)}, \beta_{k}^{(t)},L_{k}^{(t)},U_{k}^{(t)})$.
				\end{itemize}
				\item Accept $r_{k}^{(t)}$ with probability $\alpha_{r}$.
				\item Update $\bm{R}$: $r_{k}^{(t-1)} \rightarrow r_{k}^{(t)}$.
			\end{itemize}
			\}
		\end{itemize}
		\}
	\end{itemize}
	\subsection{Conditional posterior of means} \label{Conditional posterior of means}
	Express the $p_{i,\text{obs}} \times 1$ data $\bm{y}_{i,\text{obs}}$ as a linear regression problem:
	$$\bm{y}_{i,\text{obs}}=\bm{X}_{i,\text{obs}}\tilde{\bm{\mu}}+\bm{\epsilon}_{i,\text{obs}}, \ \ \ \bm{\epsilon}_{i,\text{obs}}\sim\text{MVN}(\bm{0},\bm{\Sigma}_{\text{obs}_{i},\text{obs}_{i}}), \text{ where}$$
	$$\tilde{\bm{\mu}}=\begin{bsmallmatrix}
		\mu_{1} \\
		\vdots \\
		\mu_{L} \\
	\end{bsmallmatrix}, \ \ \ \bm{X}_{i,\text{obs}}=\begin{bsmallmatrix}
		\mathds{1}\{\ell=1\} & ... & \mathds{1}\{\ell=L\} \\
		\vdots \\
		\mathds{1}\{\ell=1\} & ... & \mathds{1}\{\ell=L\}
	\end{bsmallmatrix}_{p_{i,\text{obs}} \times L}, \ \ \ \bm{\Sigma}_{\text{obs}_{i},\text{obs}_{i}} \text { is } \ p_{i,\text{obs}} \times p_{i,\text{obs}},$$
	$$p_{i,\text{obs}}=\sum_{j=1}^{J}\sum_{\ell=1}^{L} \mathds{1}\{y_{ij\ell} \text{ observed}\}.$$
	$$\Rightarrow \bm{Y}_{i,\text{obs}}|\tilde{\bm{\mu}},\bm{\Sigma}_{\text{obs}_{i},\text{obs}_{i}} \stackrel{\text{ind}}{\sim} \text{MVN}(\bm{X}_{i,\text{obs}}\tilde{\bm{\mu}}, \bm{\Sigma}_{\text{obs}_{i},\text{obs}_{i}}).$$
	$$\Rightarrow \text{E}[\bm{Y}_{i,\text{obs}}]=\bm{X}_{i,\text{obs}}\tilde{\bm{\mu}}=\bm{\mu}_{\text{obs}_{i}}=(\mu_{1}, ..., \mu_{L}, \ \ \ ... \ \ \ , \mu_{1}, ..., \mu_{L})_{p_{i,\text{obs}} \times 1}'.$$
	This allows a conjugate normal prior for $\tilde{\bm{\mu}}$. We use hyperparameters that are appropriately diffuse while scaled to the data \cite{Taddy}:
	$$\tilde{\bm{\mu}}\sim\text{MVN}(\bm{\mu}_{0},\bm{\Sigma}_{0}), \text{ where}$$
	$$\bm{\mu}_{0}=(\bar{y}_{1,{\text{obs}}},...,\bar{y}_{L,{\text{obs}}})', \ \ \ \bm{\Sigma}_{0}=\text{diag}\bigg(\bigg(\frac{\text{range}(\bm{y}_{1,{\text{obs}}})}{4}\bigg)^{2},...,\bigg(\frac{\text{range}(\bm{y}_{L,{\text{obs}}})}{4}\bigg)^{2}\bigg), $$
	$$\text{range}(\bm{y}_{\ell,{\text{obs}}})=\text{max}\{y_{ij\ell}  : y_{ij\ell} \text{ observed}\}-\text{min}\{y_{ij\ell} : y_{ij\ell} \text{ observed}\}.$$
	Given the data and the previous iteration value of the covariance matrix, sample the $L \times 1$ unique muscle means from the conditional posterior:
	$$\tilde{\bm{\mu}}^{(t)}|\bm{y}_{\text{obs}},\bm{\Sigma}^{(t-1)}$$
	$$\sim\text{MVN}\Bigg(
		\Bigg[\sum_{i=1}^{N}\bm{X}_{i,\text{obs}}'\bm{\Sigma}^{-1(t-1)}_{\text{obs}_{i},\text{obs}_{i}}\bm{X}_{i,\text{obs}}+\bm{\Sigma}_{0}^{-1}\Bigg]^{-1}\Bigg[\sum_{i=1}^{N}\bm{X}_{i,\text{obs}}'\bm{\Sigma}^{-1(t-1)}_{\text{obs}_{i},\text{obs}_{i}}\bm{y}_{i,\text{obs}}+\bm{\Sigma}_{0}^{-1}\bm{\mu}_{0}\Bigg],$$
	$$\Bigg[\sum_{i=1}^{N}\bm{X}_{i,\text{obs}}'\bm{\Sigma}^{-1(t-1)}_{\text{obs}_{i},\text{obs}_{i}}\bm{X}_{i,\text{obs}}+\bm{\Sigma}_{0}^{-1}\Bigg]^{-1} \Bigg).$$
	\subsubsection*{Deriving the conditional posterior of means}
	Conjugate prior:
	$$\tilde{\bm{\mu}}\sim\text{MVN}(\bm{\mu}_{0},\bm{\Sigma}_{0}).$$
	Conditional posterior:
	$$\therefore \tilde{\bm{\mu}}|\bm{y}_{\text{obs}},\bm{\Sigma}$$
	$$\sim\text{MVN}\Bigg(
		\Bigg[\sum_{i=1}^{N}\bm{X}_{i,\text{obs}}'\bm{\Sigma}^{-1}_{\text{obs}_{i},\text{obs}_{i}}\bm{X}_{i,\text{obs}}+\bm{\Sigma}_{0}^{-1}\Bigg]^{-1}\Bigg[\sum_{i=1}^{N}\bm{X}_{i,\text{obs}}'\bm{\Sigma}^{-1}_{\text{obs}_{i},\text{obs}_{i}}\bm{y}_{i,\text{obs}}+\bm{\Sigma}_{0}^{-1}\bm{\mu}_{0}\Bigg],$$
	$$\Bigg[\sum_{i=1}^{N}\bm{X}_{i,\text{obs}}'\bm{\Sigma}^{-1}_{\text{obs}_{i},\text{obs}_{i}}\bm{X}_{i,\text{obs}}+\bm{\Sigma}_{0}^{-1}\Bigg]^{-1} \Bigg).$$
	
	\subsection{Candidate distribution of \texorpdfstring{$\ell$}{l}th variance} \label{Candidate distribution of lth variance}
	Under the assumption of marginal data on just the $\ell$th muscle, $\bm{y}_{\ell}=(y_{11\ell}, ..., y_{1J\ell}, \ \ \ ... \ \ \ , y_{N1\ell}, ..., y_{NJ\ell})'$, where subjects and measurement times are mutually independent, we can derive the posterior from the following inverse-gamma conjugate prior for $s_{\ell}^{2}$:
	$$s_{\ell}^{2}\sim\text{Inv-Gamma}(\nu_{\ell}, (\nu_{\ell}+1)\hat{s}_{\ell,{\text{obs}}}^{2}).$$
	For our model, we require a correction through Metropolis-Hastings by using this approximate posterior as the candidate distribution. The hyperparameter $\nu_{\ell}$ should be greater than 2 such that the first and second moments of the prior are defined. However, we want to keep the hyperparameter small enough to not overstate our certainty of it in the presence of an informative, data-dependent prior. Hence, we chose the hyperparameters to be $\nu_{\ell}=2.1$, $\ell=1,...,L$. Given the data and the current iteration values of the unique muscle means, the candidate distribution of the $\ell$th variance candidate at $t$th iteration is:
	$$f_{c}(s_{\ell}^{2(t)}) = \text{density-Inv-Gamma}\Bigg(s_{\ell}^{2(t)}; 2.1+\frac{NJ}{2},$$ 
	$$3.1\hat{s}_{\ell,\text{obs}}^{2}+\frac{1}{2}\sum_{i=1}^{N}\sum_{j=1}^{J}(y_{ij\ell}\mathds{1}\{y_{ij\ell} \text{ observed}\}-\mu_{\ell}^{(t)})^{2}\Bigg).$$
	The acceptance rates will be relatively high (Table \ref{table:A0}) since the candidate distribution is a good approximation to the full conditional.
	
	\begin{table}[H]
	\centering
		\begin{tabular}{l|ccc}
			& Early ambulatory & Late ambulatory & Non-ambulatory \\
			\hline
			$s_{1}$[SOL] & 67.3\% & 80.3\% & 75.1\% \\
			$s_{2}$[VL] & 68.1\% & 80.9\% & 81.9\% \\
			$s_{3}$[BB] & 63.3\% & 74.7\% & 79.4\% \\
			$s_{4}$[DEL] & 65.1\% & 68.6\% & 71.8\%
		\end{tabular}
		\caption{Acceptance rates of standard deviations}
		\label{table:A0}
	\end{table}

	\subsection{Barnard approach details} \label{Supplement - Barnard approach details}
	Input: $i$th index, $j$th index, $\bm{R}$ matrix, where $\bm{R}$ in this section is an arbitrary correlation matrix, and the $(i,j)$th element is a unique element in the matrix.
	\\ \\
	Output: PD interval for the $(i,j)$th element of $\bm{R}$.
	\\ \\
	Procedure:
	\begin{enumerate}[nosep]
		\item Replace the $(i,j)$th element and $(j,i)$th element of $\bm{R}$ with 1. Call this augmented matrix $\bm{R}_{1}$.
		\item Replace the $(i,j)$th element and $(j,i)$th element of $\bm{R}$ with -1. Call this augmented matrix $\bm{R}_{-1}$.
		\item Replace the $(i,j)$th element and $(j,i)$th element of $\bm{R}$ with 0. Call this augmented matrix $\bm{R}_{0}$.
		\item Compute coefficients of quadratic expression, $au^{2}+bu+c$, with the following:
		$$a=\frac{|\bm{R}_{1}| + |\bm{R}_{-1}| - 2|\bm{R}_{0}|}{2},$$
		$$b=\frac{|\bm{R}_{1}|-|\bm{R}_{-1}|}{2},$$
		$$c=|\bm{R}_{0}|.$$
		\item Solve for roots $u_{1}$ and $u_{2}$ of the quadratic expression.
		\item The PD interval comes out to be
		$$(L,U)=(\min\{u_{1},u_{2}\}, \max\{u_{1},u_{2}\}).$$
	\end{enumerate}
	
	\subsection{Building the largest possible submatrices for different correlation parameters} \label{Supplement - Building the largest}
	We build the largest possible submatrix for correlation element $r_{k}$, $k=1,...,q$, by reordering the elements of the structured correlation matrix $\bm{R}$, such that the resulting submatrix has maximal dimension and contains the $r_{k}$ element exactly once. In other words, we select a subset of the original outcomes, $\hat{\bm{y}} \subset \bm{y}$, corresponding to the largest possible submatrix of $\bm{R}$ that contains the desired correlation element $\text{Corr}(Y_{m}, Y_{n})$ only once:
	$$\text{Corr}(Y_{m}, Y_{n}) \neq \text{Corr}(Y_{u}, Y_{m}),$$
	$$\text{Corr}(Y_{m}, Y_{n}) \neq \text{Corr}(Y_{u}, Y_{n}),$$
	$$\text{Corr}(Y_{m}, Y_{n}) \neq \text{Corr}(Y_{u}, Y_{v}^{*}),$$
	$\forall y_{u}, y_{v} \in \hat{\bm{y}}$. For simplicity, we reorder $\bm{R}$ such that the resulting submatrix is its leading principal submatrix. However, we note that the Barnard approach can be adapted to any principal submatrix, not just the leading one, since every principal submatrix of a PD matrix is itself PD \cite{Harville}. That is, the submatrix itself can be treated as a valid correlation matrix.
	\\ \\
	If $y_{m}$ and $y_{n}$ are outcomes that correspond to $\eta_{ll'}$, we build the largest possible submatrix as follows:
	\begin{enumerate}[nosep]
		\item Add $y_{m}$ and $y_{n}$ as the first two elements of $\hat{\bm{y}}$. If $y_{m}$ and $y_{n}$ correspond to $\eta_{\ell\ell'}$, then WLOG they must correspond to the same measurement time $j$ but to different muscles $\ell$ and $\ell'$, respectively. \label{eta step 1}
		\item Add the remaining outcomes to $\hat{\bm{y}}$ that correspond to measurement time $j$ since all pairwise combinations of outcomes that correspond to the same measurement time will produce the set of unique $\eta$ elements inside the submatrix. \label{eta step 2}
		\item Add all but one outcome to $\hat{\bm{y}}$ that correspond to arbitrary measurement time $j' \neq j$, where the one outcome we exclude corresponds to either muscle $\ell$ or $\ell'$. For instance, if we already added the outcome that corresponds to measurement time $j'$ and muscle $\ell$, we must exclude the outcome that corresponds to measurement time $j'$ and muscle $\ell'$. This ensures that $\eta_{\ell\ell'}$ does not reappear in the submatrix. \label{eta step 3}
		\item Do the previous step $\forall j' \neq j$. This exhausts our options of outcomes that we can add to $\hat{\bm{y}}$ while ensuring that $\eta_{\ell\ell'}$ remains a unique element inside the submatrix. \label{eta step 4}
	\end{enumerate}
	At the end of the procedure, $\hat{\bm{y}}$ contains $L + (L-1)(J-1)$ elements. The first term $L$ is the number of outcomes added to $\hat{\bm{y}}$ from Steps \ref{eta step 1} to \ref{eta step 2}, while the second term $(L-1)(J-1)$ is the number of outcomes added from Steps \ref{eta step 3} to \ref{eta step 4}. Hence, the largest possible submatrix for $\eta_{\ell\ell'}$ is an $[L + (L-1)(J-1)] \times [L + (L-1)(J-1)]$ matrix within the $JL \times JL$ structured correlation matrix. There are $J$ variations of the largest possible submatrix for $\eta_{\ell\ell'}$ since there are $J$ possible combinations of omitting either muscle $\ell$ or $\ell'$ across the $J-1$ time points in Steps \ref{eta step 3} and \ref{eta step 4}. We consider combinations and not permutations because the count of each unique element inside the submatrix is not affected by how we order the time points in Steps \ref{eta step 3} and \ref{eta step 4} (i.e. \textit{when} we omit muscle $\ell$ does not matter). How we order the time points in Steps \ref{eta step 3} and \ref{eta step 4} does not affect the determinant calculations of the Barnard approach \cite{Barnard} (Section \ref{Supplement - Barnard approach details}). Furthermore, note that the ordering of muscles for a particular time point does not matter for the same reason, so we do not consider this.
	\\ \\
	If $y_{m}$ and $y_{n}$ are outcomes that correspond to $\rho_{(\ell)}$, we build the largest possible submatrix as follows:
	\begin{enumerate}[nosep]
		\item Add $y_{m}$ and $y_{n}$ as the first two elements of $\hat{\bm{y}}$. If $y_{m}$ and $y_{n}$ correspond to $\rho_{(\ell)}$, then WLOG they must correspond to the same muscle $\ell$ but to different measurement times $j$ and $j'$, respectively. \label{rho step 1}
		\item Add the remaining outcomes to $\hat{\bm{y}}$ that correspond to measurement time $j$ since all pairwise combinations of outcomes that correspond to the same measurement time will produce the set of unique $\eta$ elements inside the submatrix. \label{rho step 2}
		\item Add the remaining outcomes to $\hat{\bm{y}}$ that correspond to measurement time $j'$ since all pairwise combinations of outcomes that correspond to the same measurement time will produce the set of unique $\eta$ elements inside the submatrix. Furthermore, all pairwise combinations of outcomes in which one outcome corresponds to measurement time $j$ and the other to measurement time $j'$ will produce the set of unique $\rho$ elements (if the pair of outcomes correspond to the same muscle) and numerous $\gamma$ elements (if the pair of outcomes correspond to different muscles) inside the submatrix. \label{rho step 3}
		\item Add all but one outcome to $\hat{\bm{y}}$ that correspond to arbitrary measurement time $j'' \neq j, j'$, where the one outcome we exclude corresponds to muscle $\ell$. This ensures $\rho_{(\ell)}$ does not reappear in the submatrix. \label{rho step 4}
		\item Do the previous step $\forall j'' \neq j,j'$. This exhausts our options of outcomes that we can add to $\hat{\bm{y}}$ while ensuring that $\rho_{(\ell)}$ remains a unique element inside the submatrix. \label{rho step 5}
	\end{enumerate}
	At the end of the procedure, $\hat{\bm{y}}$ contains $2L + (L-1)(J-2)$ elements. The first term $2L$ is the number of outcomes added to $\hat{\bm{y}}$ from Steps \ref{rho step 1} to \ref{rho step 3}, while the second term $(L-1)(J-2)$ is the number of outcomes added from Steps \ref{rho step 4} to \ref{rho step 5}. Hence, the largest possible submatrix for $\rho_{(\ell)}$ is a $[2L+(L-1)(J-2)] \times [2L+(L-1)(J-2)]$ matrix within the $JL \times JL$ structured correlation matrix. Note that there are no other variations of the largest possible submatrix for $\rho_{(\ell)}$.
	\\ \\
	If $y_{m}$ and $y_{n}$ are outcomes that correspond to $\gamma$, then we build the largest possible submatrix as follows:
	\begin{enumerate}[nosep]
		\item Add $y_{m}$ and $y_{n}$ as the first two elements of $\hat{\bm{y}}$. If $y_{m}$ and $y_{n}$ correspond to $\gamma$, then WLOG they must correspond to different measurement times $j$ and $j'$ as well as different muscles $\ell$ and $\ell'$, respectively. \label{gamma step 1}
		\item Add an outcome to $\hat{\bm{y}}$ that corresponds to the same muscle as $y_{m}$ and same measurement time as $y_{n}$, or add an outcome that corresponds to the same measurement time as $y_{m}$ and same muscle as $y_{n}$. This will produce $\eta_{\ell\ell'}$ and either $\rho_{(\ell)}$ or $\rho_{(\ell')}$ inside the submatrix, respectively. \label{gamma step 2}
		
	\end{enumerate}
	We cannot add anymore outcomes as this would result in another $\gamma$ element in the submatrix. If we were to add a fourth outcome to $\hat{\bm{y}}$, it must somehow satisfy the same condition described in Step \ref{gamma step 2} as when we added the third outcome. This is only possible if the fourth outcome shares neither muscle nor measurement time with the third outcome; however, in that case, the fourth outcome would inevitably introduce another $\gamma$ element into the submatrix. Therefore, the largest possible submatrix for $\gamma$ can be at most a $3 \times 3$ matrix within the $JL \times JL$ structured correlation matrix, and it must contain 3 unique elements: $\gamma$, $\eta_{\ell\ell'}$, and either $\rho_{(\ell)}$ or $\rho_{(\ell')}$. In total, there are $2{L \choose 2}$ combinations of this type of submatrix since there are ${L \choose 2}$ unique $\eta_{\ell\ell'}$ elements, and we consider either $\rho_{(\ell)}$ or $\rho_{(\ell')}$.
	\subsubsection{Algorithms} \label{Supplement - Subset Algo}
	\begin{algorithm}[H]
		
		\caption{Vector of outcomes that correspond to a largest possible submatrix for $r_{k}$}\label{general_algo}
		\KwIn{$\bm{y}=(y_{1},...,y_{p})'$}
		\KwOut{$\hat{\bm{y}} \subset \bm{y}$}
		$\hat{\bm{y}} \gets (y_{m},y_{n})'$, where $y_{m}$ and $y_{n}$ are outcomes that correspond to correlation $r_{k}$\;
		$\bm{y} \gets \bm{y}\setminus\hat{\bm{y}}$\;
		\While{$\bm{y} \neq \emptyset$}{
			Sample $y_{u}$ from $\bm{y}$\;
			\If{$r_{k}$ is a unique element inside $\bm{R}((\hat{\bm{y}}',y_{u})')$ }{
				$\hat{\bm{y}} \gets (\hat{\bm{y}}',y_{u})'$\;
				}
			$\bm{y} \gets \bm{y}\setminus\{y_{u}\}$\; 
		}
	\end{algorithm}
	\medbreak
	\begin{algorithm}[H]
		\caption{Vectors of outcomes that correspond to the largest possible submatrices for $\eta_{\ell\ell'}$}\label{eta_algo}
		\KwIn{$\bm{y}=(y_{1},...,y_{p})'$ outcomes and fixed measurement time $j$}
		\KwOut{$\hat{\bm{y}}_{c}, \ c=1,...,J$, where $\hat{\bm{y}}_{c}$ is a $(LJ-J+1) \times 1$ vector of outcomes}
		$\bm{V}=\begin{bsmallmatrix}
			\ell & \ell & \ell & ... & \ell \\
			\ell' & \ell & \ell & ... & \ell \\
			\ell' & \ell' & \ell & ... & \ell \\
			\vdots \\
			\ell' & \ell' & \ell' & ... & \ell'
		\end{bsmallmatrix}_{J \times (J-1)}$ represents the $J$ combinations of omitting either muscle $\ell$ or $\ell'$ for $J-1$ measurement times not equal to $j$\;
		\For{$c=1,...,J$}{
			$\hat{\bm{y}}_{c} \gets \bm{y}_{j}$, where $\bm{y}_{j}$ are outcomes that correspond to measurement time $j$\;
			$\bm{v} \gets \bm{V}[c,]$\;
			\For{$j' \neq j$}{
				$\hat{\bm{y}}_{c} \gets (\hat{\bm{y}}_{c}',\bm{y}_{-(\bm{v}[j'])}')'$, where $\bm{y}_{-(\bm{v}[j'])}$ are outcomes that correspond to measurement time $j'$ and muscles excluding $\bm{v}[j']$\;
			}
		}
	\end{algorithm}
	\medbreak
	\begin{algorithm}[H]
		\caption{Vector of outcomes that corresponds to the largest possible submatrix for $\rho_{(\ell)}$}\label{rho_algo}
		\KwIn{$\bm{y}=(y_{1},...,y_{p})'$ outcomes and fixed  measurement times $j$, $j'$}
		\KwOut{$\hat{\bm{y}}$, where $\hat{\bm{y}}$ is a $(LJ-J+2) \times 1$ vector of outcomes}
		$\hat{\bm{y}}\gets(\bm{y}_{j}',\bm{y}_{j'}')'$, where $\bm{y}_{j}$ and $\bm{y}_{j'}$ are outcomes that correspond to measurement times $j$ and $j'$, respectively\;
		\For{$j'' \neq j,j'$}{
			$\hat{\bm{y}} \gets (\hat{\bm{y}}',\bm{y}_{j''(-\ell)}')'$, where $\bm{y}_{j''(-\ell)}$ are outcomes that correspond to measurement time $j''$ and muscles excluding $\ell$\;
		}
	\end{algorithm}
	\medbreak
	\begin{algorithm}[H]
		\caption{Vectors of outcomes that correspond to the largest possible submatrices for $\gamma$}\label{gamma_algo}
		\KwIn{$\bm{y}=(y_{1},...,y_{p})'$ outcomes}
		\KwOut{$\hat{\bm{y}}_{c}, \ c=1,...,2{L \choose 2}$, where $\hat{\bm{y}}_{c}$ is a $3 \times 1$ vector of outcomes}
		$\bm{v}_{2{L \choose 2}\times1}=(\{\gamma,\eta_{12},\rho_{(1)}\},\{\gamma,\eta_{12},\rho_{(2)}\},...,\{\gamma,\eta_{L-1,L},\rho_{(L-1)}\},\{\gamma,\eta_{L-1,L},\rho_{(L)}\})'$\;
		\For{$c=1,...,2{L \choose 2}$}{
			$\hat{\bm{y}}_{c} \gets (y_{m},y_{n},y_{u})'$, where $y_{m}$, $y_{n}$, $y_{u}$ are outcomes that correspond to correlations $\bm{v}[c]$\;
		}
	\end{algorithm}

	\section{Supporting Information B} \label{Supplement B}
	\subsection{Supporting Information (Section \ref{Data and motivation})} \label{Supplement B -- Data and motivation}
	\begin{table}[H]
	\centering
		\begin{tabular}{l|cccc}
			& SOL missingness & VL missingness & BB missingness & DEL missingness \\
			\hline
			Early ambulatory & 5/419=1.2\% & 21/419=5.0\% & 372/419=88.8\% & 357/419=85.2\% \\
			Late ambulatory & 2/128=1.6\% & 7/128=5.5\% & 95/128=74.2\% & 94/128=73.4\% \\
			Non-ambulatory & 5/115=4.4\% & 20/115=17.4\% & 77/115=67.0\% & 69/115=60.0\%
		\end{tabular}
		\caption{Missingness by muscle (Section \ref{Data and motivation})}
		\label{table:DM1}
	\end{table}
	
	\begin{table}[H]
	\centering
		\begin{tabular}{l|ccccc}
			& 1 missing & 2 missing & 3 missing & 4 observed \\
			\hline
			Early ambulatory & 22/419=5.3\% & 332/419=79.2\% & 23/419=5.5\% & 42/419=10.0\% \\
			Late ambulatory & 12/128=9.4\% & 87/128=68.0\% & 4/128=3.1\% & 25/128=19.5\% \\
			Non-ambulatory & 17/115=14.8\% & 53/115=46.1\% & 16/115=13.9\% & 29/115=25.2\%
		\end{tabular}
		\caption{Distribution of the number of missing muscle measurements at a given measurement time (Section \ref{Data and motivation})}
		\label{table:DM2}
	\end{table}
	
	\begin{table}[H]
	\centering
		\begin{tabular}{l|cccccccc}
			& $J=1$ & $J=2$ & $J=3$ & $J=4$ & $J=5$ & $J=6$ & $J=7$ & $J=8$ \\
			\hline
			Early ambulatory & 36/140 & 26/140 & 31/140 & 23/140 & 9/140 & 7/140 & 5/140 & 3/140 \\
			Late ambulatory & 40/74 & 19/74 & 12/74 & 2/74 & 0/74 & 1/74 & NA & NA \\
			Non-ambulatory & 20/51 & 12/51 & 9/51 & 8/51 & 1/51 & 0/51 & 1/51 & NA \\
		\end{tabular}
		\caption{Distribution of the number of subject measurement times (Section \ref{Data and motivation})}
		\label{table:DM3}
	\end{table}
	
	\subsection{Supporting Information (Section \ref{Simulations})} \label{Supplement B -- Simulations}
	\begin{table}[H]
    \centering
        \begin{tabular}{l|ccccc}
        Root MSE & $w_{1}$ & $w_{2}$ & $w_{3}$ & $w_{4}$ & SRM \\
        \hline
        Normal -- no miss -- early ambulatory & 0.09 & 0.08 & 0.06 & 0.00 & 0.08 \\
        Normal -- no miss -- late ambulatory & 0.05 & 0.04 & 0.05 & 0.07 & 0.08 \\
        Normal -- no miss -- non-ambulatory & 0.05 & 0.03 & 0.04 & 0.02 & 0.07 \\
        \hline
        $t_{10}$ -- no miss -- early ambulatory & 0.10 & 0.08 & 0.06 & 0.00 & 0.09 \\
        $t_{10}$ -- no miss -- late ambulatory & 0.05 & 0.04 & 0.06 & 0.07 & 0.09 \\
        $t_{10}$ -- no miss -- non-ambulatory & 0.05 & 0.04 & 0.05 & 0.03 & 0.08 \\
        \hline
        $t_{3}$ -- no miss -- early ambulatory & 0.14 & 0.14 & 0.09 & 0.03 & 0.20 \\
        $t_{3}$ -- no miss -- late ambulatory & 0.09 & 0.07 & 0.09 & 0.09 & 0.20 \\
        $t_{3}$ -- no miss -- non-ambulatory & 0.09 & 0.05 & 0.09 & 0.05 & 0.20 \\
        \hline
        Skew-normal -- no miss -- early ambulatory & 0.09 & 0.10 & 0.05 & 0.02 & 0.27 \\
        Skew-normal -- no miss -- late ambulatory & 0.05 & 0.04 & 0.05 & 0.07 & 0.23 \\
        Skew-normal -- no miss -- non-ambulatory & 0.05 & 0.04 & 0.06 & 0.04 & 0.16 \\
        \hline
        Normal -- yes miss -- early ambulatory & 0.10 & 0.10 & 0.09 & 0.03 & 0.09 \\
        Normal -- yes miss -- late ambulatory & 0.07 & 0.05 & 0.09 & 0.10 & 0.09 \\
        Normal -- yes miss -- non-ambulatory & 0.07 & 0.04 & 0.07 & 0.05 & 0.11 \\
        \end{tabular}
        \caption{Root MSEs of weights and SRM (Section \ref{Coverage, bias, MSE of weights and SRM})}
		\label{table:sim3_extra}
    \end{table}
    
    \begin{table}[H]
			\centering
			\begin{adjustbox}{max width=\textwidth}
			\begin{tabular}{l|cccccc}
				Early & $\text{R-Beta}(L,U)$ & $\text{R-Beta}(L_{1},U_{1})$ & $\text{R-Beta}(-1,1)$ & $\text{Unif}(L,U)$ & $\text{Unif}(L_{1},U_{1})$ & $\text{Unif}(-1,1)$ \\
				\hline
				$\eta_{12}$ & $\approx$ 25\% & $\approx$ 25\% & $\approx$ 25\% & 5.6\% & 5.5\% & 4.2\% \\
				$\eta_{13}$ & $\approx$ 25\% & $\approx$ 25\% & $\approx$ 25\% & 7.7\% & 7.6\% & 4.7\% \\
				$\eta_{14}$ & $\approx$ 25\% & $\approx$ 25\% & $\approx$ 25\% & 7.5\% & 7.3\% & 4.6\% \\
				$\eta_{23}$ & $\approx$ 25\% & $\approx$ 25\% & $\approx$ 25\% & 7.9\% & 7.6\% & 4.8\% \\
				$\eta_{24}$ & $\approx$ 25\% & $\approx$ 25\% & $\approx$ 25\% & 7.7\% & 7.6\% & 4.8\% \\
				$\eta_{34}$ & $\approx$ 25\% & $\approx$ 25\% & $\approx$ 25\% & 6.6\% & 6.3\% & 4.6\% \\
				$\rho_{(1)}$ & $\approx$ 25\% & --- & $\approx$ 25\% & 6.3\% & --- & 3.7\% \\
				$\rho_{(2)}$ & $\approx$ 25\% & --- & $\approx$ 25\% & 6.4\% & --- & 3.8\% \\
				$\rho_{(3)}$ & $\approx$ 25\% & --- & $\approx$ 25\% & 7.5\% & --- & 5.4\% \\
				$\rho_{(4)}$ & $\approx$ 25\% & --- & $\approx$ 25\% & 7.3\% & --- & 4.7\% \\
				$\gamma$ & $\approx$ 25\% & $\approx$ 25\% & $\approx$ 25\% & 2.2\% & 2.0\% & 1.7\% \\
			\end{tabular}
			\end{adjustbox}
			\caption{Acceptance rates of correlations for early ambulatory data (Section
			\ref{Acceptance and positive definiteness rates of correlations})}
			\captionsetup{font=footnotesize}
			\caption*{\footnotesize \emph{Note:} $(L, U)$=$(L_{1}, U_{1})$ for $\rho_{(\ell)}$, i.e. only one variation of largest possible submatrix for $\rho_{(\ell)}$}
			\label{table:sim4}
		\end{table}
	
	\begin{table}[H]
		\centering
		\begin{adjustbox}{max width=\textwidth}
		\begin{tabular}{l|cccccc}
			Late & $\text{R-Beta}(L,U)$ & $\text{R-Beta}(L_{1},U_{1})$ & $\text{R-Beta}(-1,1)$ & $\text{Unif}(L,U)$ & $\text{Unif}(L_{1},U_{1})$ & $\text{Unif}(-1,1)$ \\
			\hline
			$\eta_{12}$ & $\approx$ 25\% & $\approx$ 25\% & $\approx$ 25\% & 7.1\% & 7.0\% & 5.6\% \\
			$\eta_{13}$ & $\approx$ 25\% & $\approx$ 25\% & $\approx$ 25\% & 7.8\% & 7.8\% & 5.8\% \\
			$\eta_{14}$ & $\approx$ 25\% & $\approx$ 25\% & $\approx$ 25\% & 7.1\% & 6.9\% & 4.8\% \\
			$\eta_{23}$ & $\approx$ 25\% & $\approx$ 25\% & $\approx$ 25\% & 7.9\% & 7.8\% & 5.7\% \\
			$\eta_{24}$ & $\approx$ 25\% & $\approx$ 25\% & $\approx$ 25\% & 7.2\% & 7.1\% & 4.9\% \\
			$\eta_{34}$ & $\approx$ 25\% & $\approx$ 25\% & $\approx$ 25\% & 5.4\% & 5.2\% & 4.1\% \\
			$\rho_{(1)}$ & $\approx$ 25\% & --- & $\approx$ 25\% & 6.1\% & --- & 4.8\% \\
			$\rho_{(2)}$ & $\approx$ 25\% & --- & $\approx$ 25\% & 4.5\% & --- & 3.6\% \\
			$\rho_{(3)}$ & $\approx$ 25\% & --- & $\approx$ 25\% & 7.1\% & --- & 4.9\% \\
			$\rho_{(4)}$ & $\approx$ 25\% & --- & $\approx$ 25\% & 7.6\% & --- & 4.6\% \\
			$\gamma$ & $\approx$ 25\% & $\approx$ 25\% & $\approx$ 25\% & 2.3\% & 2.0\% & 1.8\% \\
		\end{tabular}
		\end{adjustbox}
		\caption{Acceptance rates of correlations for late ambulatory data (Section \ref{Acceptance and positive definiteness rates of correlations})}
		\captionsetup{font=footnotesize}
		\caption*{\footnotesize \emph{Note:} $(L, U)$=$(L_{1}, U_{1})$ for $\rho_{(\ell)}$, i.e. only one variation of largest possible submatrix for $\rho_{(\ell)}$}
		\label{table:sim5}
	\end{table}
	
	\begin{table}[H]
		\centering
		\begin{adjustbox}{max width=\textwidth}
		\begin{tabular}{l|cccccc}
			Non & $\text{R-Beta}(L,U)$ & $\text{R-Beta}(L_{1},U_{1})$ & $\text{R-Beta}(-1,1)$ & $\text{Unif}(L,U)$ & $\text{Unif}(L_{1},U_{1})$ & $\text{Unif}(-1,1)$ \\
			\hline
			$\eta_{12}$ & $\approx$ 25\% & $\approx$ 25\% & $\approx$ 25\% & 6.1\% & 6.1\% & 5.2\% \\
			$\eta_{13}$ & $\approx$ 25\% & $\approx$ 25\% & $\approx$ 25\% & 7.8\% & 7.8\% & 5.6\% \\
			$\eta_{14}$ & $\approx$ 25\% & $\approx$ 25\% & $\approx$ 25\% & 5.4\% & 5.3\% & 4.0\% \\
			$\eta_{23}$ & $\approx$ 25\% & $\approx$ 25\% & $\approx$ 25\% & 7.9\% & 7.9\% & 6.4\% \\
			$\eta_{24}$ & $\approx$ 25\% & $\approx$ 25\% & $\approx$ 25\% & 8.0\% & 7.6\% & 5.3\% \\
			$\eta_{34}$ & $\approx$ 25\% & $\approx$ 25\% & $\approx$ 25\% & 5.9\% & 5.7\% & 4.6\% \\
			$\rho_{(1)}$ & $\approx$ 25\% & --- & $\approx$ 25\% & 6.6\% & --- & 4.2\% \\
			$\rho_{(2)}$ & $\approx$ 25\% & --- & $\approx$ 25\% & 5.3\% & --- & 4.4\% \\
			$\rho_{(3)}$ & $\approx$ 25\% & --- & $\approx$ 25\% & 7.2\% & --- & 5.8\% \\
			$\rho_{(4)}$ & $\approx$ 25\% & --- & $\approx$ 25\% & 7.3\% & --- & 4.9\% \\
			$\gamma$ & $\approx$ 25\% & $\approx$ 25\% & $\approx$ 25\% & 2.3\% & 2.0\% & 1.9\% \\
		\end{tabular}
		\end{adjustbox}
		\caption{Acceptance rates of correlations for non-ambulatory data (Section \ref{Acceptance and positive definiteness rates of correlations})}
		\captionsetup{font=footnotesize}
		\caption*{\footnotesize \emph{Note:} $(L, U)$=$(L_{1}, U_{1})$ for $\rho_{(\ell)}$, i.e. only one variation of largest possible submatrix for $\rho_{(\ell)}$}
		\label{table:sim6}
	\end{table}
	
	\begin{table}[H]
		\centering
		\begin{adjustbox}{max width=\textwidth}
		\begin{tabular}{l|cccccc}
			Early & $\text{R-Beta}(L,U)$ & $\text{R-Beta}(L_{1},U_{1})$ & $\text{R-Beta}(-1,1)$ & $\text{Unif}(L,U)$ & $\text{Unif}(L_{1},U_{1})$ & $\text{Unif}(-1,1)$ \\
			\hline
			$\eta_{12}$ & 100\% & 100\% & 100\% & 90.6\% & 90.0\% & 67.4\% \\
			$\eta_{13}$ & 100\% & 100\% & 100\% & 92.8\% & 89.9\% & 56.2\% \\
			$\eta_{14}$ & 100\% & 100\% & 100\% & 94.0\% & 93.4\% & 60.1\% \\
			$\eta_{23}$ & 100\% & 100\% & 100\% & 92.1\% & 88.9\% & 55.4\% \\
			$\eta_{24}$ & 100\% & 100\% & 100\% & 94.5\% & 93.8\% & 59.1\% \\
			$\eta_{34}$ & 100\% & 100\% & 100\% & 88.2\% & 85.7\% & 62.4\% \\
			$\rho_{(1)}$ & 96.5\% & --- & 95.7\% & 66.7\% & --- & 39.1\% \\
			$\rho_{(2)}$ & 95.6\% & --- & 94.5\% & 66.7\% & --- & 39.7\% \\
			$\rho_{(3)}$ & 99.4\% & --- & 98.9\% & 66.4\% & --- & 47.9\% \\
			$\rho_{(4)}$ & 96.9\% & --- & 96.2\% & 66.6\% & --- & 43.2\% \\
			$\gamma$ & 98.8\% & 98.8\% & 98.8\% & 32.9\% & 28.8\% & 24.9\% \\
		\end{tabular}
		\end{adjustbox}
		\caption{Positive definiteness rates of correlations for early ambulatory data (Section \ref{Acceptance and positive definiteness rates of correlations})}
		\captionsetup{font=footnotesize}
		\caption*{\footnotesize \emph{Note:} $(L, U)$=$(L_{1}, U_{1})$ for $\rho_{(\ell)}$, i.e. only one variation of largest possible submatrix for $\rho_{(\ell)}$}
		\label{table:sim7}
	\end{table}
	
	\begin{table}[H]
		\centering
		\begin{adjustbox}{max width=\textwidth}
		\begin{tabular}{l|cccccc}
			Late & $\text{R-Beta}(L,U)$ & $\text{R-Beta}(L_{1},U_{1})$ & $\text{R-Beta}(-1,1)$ & $\text{Unif}(L,U)$ & $\text{Unif}(L_{1},U_{1})$ & $\text{Unif}(-1,1)$ \\
			\hline
			$\eta_{12}$ & 100\% & 100\% & 100\% & 84.1\% & 82.6\% & 67.8\% \\
			$\eta_{13}$ & 100\% & 100\% & 100\% & 92.0\% & 91.2\% & 67.3\% \\
			$\eta_{14}$ & 100\% & 100\% & 100\% & 94.1\% & 91.7\% & 64.7\% \\
			$\eta_{23}$ & 100\% & 100\% & 100\% & 86.6\% & 84.5\% & 62.3\% \\
			$\eta_{24}$ & 100\% & 100\% & 100\% & 96.8\% & 94.7\% & 66.4\% \\
			$\eta_{34}$ & 100\% & 100\% & 100\% & 91.8\% & 90.0\% & 71.2\% \\
			$\rho_{(1)}$ & 94.4\% & --- & 92.4\% & 66.7\% & --- & 52.1\% \\
			$\rho_{(2)}$ & 94.1\% & --- & 93.0\% & 66.7\% & --- & 53.4\% \\
			$\rho_{(3)}$ & 97.8\% & --- & 97.1\% & 66.4\% & --- & 46.2\% \\
			$\rho_{(4)}$ & 99.2\% & --- & 98.6\% & 66.6\% & --- & 40.2\% \\
			$\gamma$ & 98.6\% & 98.6\% & 98.6\% & 28.4\% & 24.1\% & 22.1\% \\
		\end{tabular}
		\end{adjustbox}
		\caption{Positive definiteness rates of correlations for late ambulatory data (Section \ref{Acceptance and positive definiteness rates of correlations})}
		\captionsetup{font=footnotesize}
		\caption*{\footnotesize \emph{Note:} $(L, U)$=$(L_{1}, U_{1})$ for $\rho_{(\ell)}$, i.e. only one variation of largest possible submatrix for $\rho_{(\ell)}$}
		\label{table:sim8}
	\end{table}
	
	\begin{table}[H]
		\centering
		\begin{adjustbox}{max width=\textwidth}
		\begin{tabular}{l|cccccc}
			Non & $\text{R-Beta}(L,U)$ & $\text{R-Beta}(L_{1},U_{1})$ & $\text{R-Beta}(-1,1)$ & $\text{Unif}(L,U)$ & $\text{Unif}(L_{1},U_{1})$ & $\text{Unif}(-1,1)$ \\
			\hline
			$\eta_{12}$ & 100\% & 100\% & 100\% & 92.9\% & 92.4\% & 78.3\% \\
			$\eta_{13}$ & 100\% & 100\% & 100\% & 93.3\% & 92.7\% & 66.1\% \\
			$\eta_{14}$ & 100\% & 100\% & 100\% & 93.1\% & 90.0\% & 69.5\% \\
			$\eta_{23}$ & 100\% & 100\% & 100\% & 94.0\% & 93.2\% & 75.2\% \\
			$\eta_{24}$ & 100\% & 100\% & 100\% & 94.7\% & 90.8\% & 63.4\% \\
			$\eta_{34}$ & 100\% & 100\% & 100\% & 89.7\% & 86.8\% & 70.2\% \\
			$\rho_{(1)}$ & 95.2\% & --- & 93.9\% & 66.7\% & --- & 42.2\% \\
			$\rho_{(2)}$ & 94.6\% & --- & 94.5\% & 66.7\% & --- & 55.1\% \\
			$\rho_{(3)}$ & 97.4\% & --- & 97.3\% & 66.4\% & --- & 53.4\% \\
			$\rho_{(4)}$ & 99.7\% & --- & 99.4\% & 66.6\% & --- & 44.6\% \\
			$\gamma$ & 99.1\% & 98.9\% & 98.9\% & 23.0\% & 19.8\% & 18.5\% \\
		\end{tabular}
		\end{adjustbox}
		\caption{Positive definiteness rates of correlations for non-ambulatory data (Section \ref{Acceptance and positive definiteness rates of correlations})}
		\captionsetup{font=footnotesize}
		\caption*{\footnotesize \emph{Note:} $(L, U)$=$(L_{1}, U_{1})$ for $\rho_{(\ell)}$, i.e. only one variation of largest possible submatrix for $\rho_{(\ell)}$}
		\label{table:sim9}
	\end{table}
		
	\subsection{Supporting Information (Section \ref{Data analysis})} \label{Supplement B -- Data analysis}
	
	\begin{table}[H]
	\centering
		\begin{tabular}{l|ccc}
			& Early ambulatory & Late ambulatory & Non-ambulatory \\
			\hline
			$\eta_{12}$[SOL,VL] & 25.4\% & 25.4\% & 25.4\% \\
			$\eta_{13}$[SOL,BB] & 25.3\% & 25.2\% & 25.6\% \\
			$\eta_{14}$[SOL,DEL] & 25.3\% & 25.5\% & 25.5\% \\
			$\eta_{23}$[VL,BB] & 25.5\% & 25.4\% & 25.3\% \\
			$\eta_{24}$[VL,DEL] & 25.5\% & 25.4\% & 25.5\% \\
			$\eta_{34}$[BB,DEL] & 25.3\% & 25.1\% & 25.5\% \\
			$\rho_{(1)}$[SOL] & 25.5\% & 25.3\% & 25.4\% \\
			$\rho_{(2)}$[VL] & 25.6\% & 25.1\% & 25.4\% \\
			$\rho_{(3)}$[BB] & 25.3\% & 25.4\% & 25.4\% \\
			$\rho_{(4)}$[DEL] & 25.4\% & 25.4\% & 25.5\% \\
			$\gamma$ & 25.1\% & 25.4\% & 25.4\% \\
		\end{tabular}
		\caption{Acceptance rates of correlations drawn from $\text{R-Beta}(L,U)$ (Section \ref{Data analysis - Computations})}
		\label{table:DA1}
	\end{table}
	
	\begin{table}[H]
	\centering
		\begin{tabular}{l|ccc}
			& Early ambulatory & Late ambulatory & Non-ambulatory \\
			\hline
			$\eta_{12}$[SOL,VL] & 6.0\% & 12.7\% & 10.9\% \\
			$\eta_{13}$[SOL,BB] & 8.0\% & 15.5\% & 15.3\% \\
			$\eta_{14}$[SOL,DEL] & 7.5\% & 14.8\% & 11.0\% \\
			$\eta_{23}$[VL,BB] & 8.2\% & 16.2\% & 15.8\% \\
			$\eta_{24}$[VL,DEL] & 7.8\% & 15.0\% & 16.0\% \\
			$\eta_{34}$[BB,DEL] & 6.6\% & 10.4\% & 11.8\% \\
			$\rho_{(1)}$[SOL] & 5.9\% & 11.1\% & 9.9\% \\
			$\rho_{(2)}$[VL] & 6.6\% & 5.0\% & 6.6\% \\
			$\rho_{(3)}$[BB] & 6.4\% & 13.9\% & 13.6\% \\
			$\rho_{(4)}$[DEL] & 6.1\% & 14.4\% & 14.8\% \\
			$\gamma$ & 1.9\% & 2.4\% & 3.2\% \\
		\end{tabular}
		\caption{Acceptance rates of correlations drawn from $\text{Unif}(L,U)$ (Section \ref{Data analysis - Computations})}
		\label{table:DA2}
	\end{table}
	
	\begin{table}[H]
	\centering
		\begin{tabular}{l|ccc}
			& Early ambulatory & Late ambulatory & Non-ambulatory \\
			\hline
			$\eta_{12}$[SOL,VL] & 3.8\% & 7.9\% & 8.0\% \\
			$\eta_{13}$[SOL,BB] & 4.0\% & 8.4\% & 9.2\% \\
			$\eta_{14}$[SOL,DEL] & 4.1\% & 7.4\% & 7.1\% \\
			$\eta_{23}$[VL,BB] & 4.0\% & 8.0\% & 10.1\% \\
			$\eta_{24}$[VL,DEL] & 4.2\% & 7.1\% & 8.8\% \\
			$\eta_{34}$[BB,DEL] & 3.8\% & 6.4\% & 8.0\% \\
			$\rho_{(1)}$[SOL] & 3.2\% & 7.3\% & 5.7\% \\
			$\rho_{(2)}$[VL] & 3.7\% & 3.5\% & 5.0\% \\
			$\rho_{(3)}$[BB] & 3.9\% & 7.7\% & 9.9\% \\
			$\rho_{(4)}$[DEL] & 3.7\% & 7.4\% & 8.8\% \\
			$\gamma$ & 1.3\% & 1.8\% & 2.5\% \\
		\end{tabular}
		\caption{Acceptance rates of correlations drawn from $\text{Unif}(-1,1)$ (Section \ref{Data analysis - Computations})}
		\label{table:DA3}
	\end{table}
	
	\begin{table}[H]
	\centering
		\begin{tabular}{l|ccc}
			Early ambulatory & $\text{R-Beta}(L,U)$ & $\text{Unif}(L,U)$ & $\text{Unif}(-1,1)$ \\
			\hline
			$\eta_{12}$[SOL,VL] & 95.7\% & 80.6\% & 53.3\% \\
			$\eta_{13}$[SOL,BB] & 97.8\% & 86.3\% & 43.6\% \\
			$\eta_{14}$[SOL,DEL] & 95.8\% & 80.5\% & 45.3\% \\
			$\eta_{23}$[VL,BB] & 97.9\% & 87.0\% & 43.4\% \\
			$\eta_{24}$[VL,DEL] & 96.4\% & 80.0\% & 44.5\% \\
			$\eta_{34}$[BB,DEL] & 97.6\% & 89.9\% & 51.4\% \\
			$\rho_{(1)}$[SOL] & 71.6\% & 57.2\% & 30.3\% \\
			$\rho_{(2)}$[VL] & 71.2\% & 57.1\% & 31.6\% \\
			$\rho_{(3)}$[BB] & 88.5\% & 57.2\% & 36.9\% \\
			$\rho_{(4)}$[DEL] & 80.3\% & 57.1\% & 34.3\% \\
			$\gamma$ & 74.4\% & 24.9\% & 17.4\% \\
		\end{tabular}
		\caption{Positive definiteness rates of correlations for early ambulatory data (Section \ref{Data analysis - Computations})}
		\label{table:DA4}
	\end{table}
	
	\begin{table}[H]
	\centering
		\begin{tabular}{l|ccc}
			Late ambulatory & $\text{R-Beta}(L,U)$ & $\text{Unif}(L,U)$ & $\text{Unif}(-1,1)$ \\
			\hline
			$\eta_{12}$[SOL,VL] & 73.9\% & 62.1\% & 39.8\% \\
			$\eta_{13}$[SOL,BB] & 83.4\% & 71.1\% & 40.2\% \\
			$\eta_{14}$[SOL,DEL] & 89.9\% & 79.1\% & 40.6\% \\
			$\eta_{23}$[VL,BB] & 75.6\% & 63.2\% & 33.0\% \\
			$\eta_{24}$[VL,DEL] & 87.3\% & 78.0\% & 38.1\% \\
			$\eta_{34}$[BB,DEL] & 94.3\% & 82.3\% & 51.4\% \\
			$\rho_{(1)}$[SOL] & 65.3\% & 60.0\% & 39.4\% \\
			$\rho_{(2)}$[VL] & 64.5\% & 60.0\% & 43.0\% \\
			$\rho_{(3)}$[BB] & 69.7\% & 59.9\% & 33.9\% \\
			$\rho_{(4)}$[DEL] & 76.6\% & 60.0\% & 30.5\% \\
			$\gamma$ & 64.1\% & 15.6\% & 11.5\% \\
		\end{tabular}
		\caption{Positive definiteness rates of correlations for late ambulatory data (Section \ref{Data analysis - Computations})}
		\label{table:DA5}
	\end{table}
	
	\begin{table}[H]
	\centering
		\begin{tabular}{l|ccc}
			Non-ambulatory & $\text{R-Beta}(L,U)$ & $\text{Unif}(L,U)$ & $\text{Unif}(-1,1)$ \\
			\hline
			$\eta_{12}$[SOL,VL] & 85.7\% & 75.1\% & 56.4\% \\
			$\eta_{13}$[SOL,BB] & 90.3\% & 79.3\% & 48.6\% \\
			$\eta_{14}$[SOL,DEL] & 97.8\% & 90.6\% & 59.6\% \\
			$\eta_{23}$[VL,BB] & 84.9\% & 72.6\% & 49.0\% \\
			$\eta_{24}$[VL,DEL] & 95.3\% & 88.7\% & 50.2\% \\
			$\eta_{34}$[BB,DEL] & 98.0\% & 90.1\% & 61.7\% \\
			$\rho_{(1)}$[SOL] & 67.5\% & 58.3\% & 33.1\% \\
			$\rho_{(2)}$[VL] & 66.0\% & 58.3\% & 43.8\% \\
			$\rho_{(3)}$[BB] & 68.6\% & 58.3\% & 42.0\% \\
			$\rho_{(4)}$[DEL] & 78.2\% & 58.3\% & 35.1\% \\
			$\gamma$ & 57.2\% & 10.2\% & 7.8\% \\
		\end{tabular}
		\caption{Positive definiteness rates of correlations for non-ambulatory data (Section \ref{Data analysis - Computations})}
		\label{table:DA6}
	\end{table}
	
	
	
	
	\begin{table}[H]
    \centering
  		\begin{tabular}{l|ccc}
  		& Early ambulatory & Late ambulatory & Non-ambulatory \\
    	\hline
    	$\mu_{1}$[SOL] & 0.026 & 0.062 & 0.047 \\
    	$\mu_{2}$[VL] & 0.066 & 0.074 & 0.032 \\
    	$\mu_{3}$[BB] & 0.036 & 0.048 & 0.075 \\
    	$\mu_{4}$[DEL] & 0.015 & 0.050 & 0.050 \\
    	\hline
    	$s_{1}$[SOL] & 0.037 & 0.064 & 0.067 \\
    	$s_{2}$[VL] & 0.073 & 0.092 & 0.097 \\
    	$s_{3}$[BB] & 0.064 & 0.070 & 0.074 \\
    	$s_{4}$[DEL] & 0.041 & 0.065 & 0.088 \\
    	\hline
    	$\eta_{12}$[SOL,VL] & 0.596 & 0.303 & 0.395 \\
		$\eta_{13}$[SOL,BB] & 0.225 & 0.201 & 0.182 \\
		$\eta_{14}$[SOL,DEL] & 0.423 & 0.379 & 0.466 \\
		$\eta_{23}$[VL,BB] & 0.275 & 0.202 & 0.039 \\
		$\eta_{24}$[VL,DEL] & 0.414 & 0.372 & 0.119 \\
		$\eta_{34}$[BB,DEL] & 0.464 & 0.523 & 0.405 \\
		$\rho_{(1)}$[SOL] & 0.239 & 0.080 & 0.007 \\
		$\rho_{(2)}$[VL] & 0.251 & -0.012 & -0.054 \\
		$\rho_{(3)}$[BB] & 0.415 & 0.111 & 0.047 \\
		$\rho_{(4)}$[DEL] & 0.286 & 0.297 & 0.247 \\
		$\gamma$ & 0.264 & 0.123 & 0.019 \\
  		\end{tabular}
  		\caption{Posterior medians of means, standard deviations, and correlations (Section \ref{Posteriors of model parameters})}
		\label{table:DA10}
	\end{table}
	
	\begin{table}[H]
	\centering
		\begin{tabular}{l|ccc}
			Early ambulatory & 2.5th & Median & 97.5th \\
			\hline
			$\mu_{1}$[SOL] & 0.023 & 0.026 & 0.030 \\
			$\mu_{2}$[VL] & 0.059 & 0.066 & 0.073 \\
			$\mu_{3}$[BB] & 0.020 & 0.036 & 0.053 \\
			$\mu_{4}$[DEL] & 0.006 & 0.015 & 0.024
		\end{tabular}
		\caption{95\% credible intervals and posterior medians of means for early ambulatory data (Section \ref{Posteriors of model parameters})}
		\label{table:A2}
	\end{table}
	\begin{table}[H]
	\centering
		\begin{tabular}{l|ccc}
			Late ambulatory & 2.5th & Median & 97.5th \\
			\hline
			$\mu_{1}$[SOL] & 0.051 & 0.062 & 0.072 \\
			$\mu_{2}$[VL] & 0.061 & 0.074 & 0.088 \\
			$\mu_{3}$[BB] & 0.0 24 & 0.048 & 0.072 \\
			$\mu_{4}$[DEL] & 0.029 & 0.050 & 0.073
		\end{tabular}
		\caption{95\% credible intervals and posterior medians of means for late ambulatory data (Section \ref{Posteriors of model parameters})}
		\label{table:A3}
	\end{table}
	\begin{table}[H]
	\centering
		\begin{tabular}{l|ccc}
			Non-ambulatory & 2.5th & Median & 97.5th \\
			\hline
			$\mu_{1}$[SOL] & 0.035 & 0.047 & 0.060 \\
			$\mu_{2}$[VL] & 0.016 & 0.032 & 0.049 \\
			$\mu_{3}$[BB] & 0.053 & 0.075 & 0.097 \\
			$\mu_{4}$[DEL] & 0.027 & 0.050 & 0.072
		\end{tabular}
		\caption{95\% credible intervals and posterior medians of means for non-ambulatory data (Section \ref{Posteriors of model parameters})}
		\label{table:A4}
	\end{table}
	
	\begin{table}[H]
	\centering
		\begin{tabular}{l|ccc}
			Early ambulatory & 2.5th & Median & 97.5th \\
			\hline
			$s_{1}$[SOL] & 0.035 & 0.037 & 0.040 \\
			$s_{2}$[VL] & 0.068 & 0.073 & 0.078 \\
			$s_{3}$[BB] & 0.053 & 0.064 & 0.080 \\
			$s_{4}$[DEL] & 0.034 & 0.041 & 0.050
		\end{tabular}
		\caption{95\% credible intervals and posterior medians of standard deviations for early ambulatory data (Section \ref{Posteriors of model parameters})}
		\label{table:A5}
	\end{table}
	\begin{table}[H]
	\centering
		\begin{tabular}{l|ccc}
			Late ambulatory & 2.5th & Median & 97.5th \\
			\hline
			$s_{1}$[SOL] & 0.057 & 0.064 & 0.073 \\
			$s_{2}$[VL] & 0.081 & 0.092 & 0.105 \\
			$s_{3}$[BB] & 0.056 & 0.070 & 0.091 \\
			$s_{4}$[DEL] & 0.051 & 0.065 & 0.086
		\end{tabular}
		\caption{95\% credible intervals and posterior medians of standard deviations for late ambulatory data (Section \ref{Posteriors of model parameters})}
		\label{table:A6}
	\end{table}
	\begin{table}[H]
	\centering
		\begin{tabular}{l|ccc}
			Non-ambulatory & 2.5th & Median & 97.5th \\
			\hline
			$s_{1}$[SOL] & 0.059 & 0.067 & 0.077 \\
			$s_{2}$[VL] & 0.084 & 0.097 & 0.112 \\
			$s_{3}$[BB] & 0.060 & 0.074 & 0.096 \\
			$s_{4}$[DEL] & 0.073 & 0.088 & 0.109
		\end{tabular}
		\caption{95\% credible intervals and posterior medians of standard deviations for non-ambulatory data (Section \ref{Posteriors of model parameters})}
		\label{table:A7}
	\end{table}
	
	\begin{table}[H]
	\centering
		\begin{tabular}{l|ccc}
			Early ambulatory & 2.5th & Median & 97.5th \\
			\hline
			$\eta_{12}$[SOL,VL] & 0.526 & 0.596 & 0.658 \\
			$\eta_{13}$[SOL,BB] & -0.046 & 0.225 & 0.481 \\
			$\eta_{14}$[SOL,DEL] & 0.180 & 0.423 & 0.605 \\
			$\eta_{23}$[VL,BB] & 0.015 & 0.275 & 0.507 \\
			$\eta_{24}$[VL,DEL] & 0.192 & 0.414 & 0.584 \\
			$\eta_{34}$[BB,DEL] & 0.194 & 0.464 & 0.662 \\
			$\rho_{(1)}$[SOL] & 0.139 & 0.239 & 0.345 \\
			$\rho_{(2)}$[VL] & 0.139 & 0.251 & 0.368 \\
			$\rho_{(3)}$[BB] & 0.126 & 0.415 & 0.835 \\
			$\rho_{(4)}$[DEL] & 0.109 & 0.286 & 0.570 \\
			$\gamma$ & 0.175 & 0.264 & 0.359 \\
		\end{tabular}
		\caption{95\% credible intervals and posterior medians of correlations for early ambulatory data (Section \ref{Posteriors of model parameters})}
		\label{table:DA7}
	\end{table}
	\begin{table}[H]
	\centering
		\begin{tabular}{l|ccc}
			Late ambulatory & 2.5th & Median & 97.5th \\
			\hline
			$\eta_{12}$[SOL,VL] & 0.128 & 0.303 & 0.456 \\
			$\eta_{13}$[SOL,BB] & -0.201 & 0.201 & 0.533 \\
			$\eta_{14}$[SOL,DEL] & 0.032 & 0.379 & 0.635 \\
			$\eta_{23}$[VL,BB] & -0.096 & 0.202 & 0.459 \\
			$\eta_{24}$[VL,DEL] & 0.094 & 0.372 & 0.580 \\
			$\eta_{34}$[BB,DEL] & 0.148 & 0.523 & 0.746 \\
			$\rho_{(1)}$[SOL] & -0.087 & 0.080 & 0.307 \\
			$\rho_{(2)}$[VL] & -0.140 & -0.012 & 0.164 \\
			$\rho_{(3)}$[BB] & -0.083 & 0.111 & 0.389 \\
			$\rho_{(4)}$[DEL] & 0.002 & 0.297 & 0.632 \\
			$\gamma$ & 0.002 & 0.123 & 0.258 \\
		\end{tabular}
		\caption{95\% credible intervals and posterior medians of correlations for late ambulatory data (Section \ref{Posteriors of model parameters})}
		\label{table:DA8}
	\end{table}
	\begin{table}[H]
	\centering
		\begin{tabular}{l|ccc}
			Non-ambulatory & 2.5th & Median & 97.5th \\
			\hline
			$\eta_{12}$[SOL,VL] & 0.203 & 0.395 & 0.553 \\
			$\eta_{13}$[SOL,BB] & -0.115 & 0.182 & 0.442 \\
			$\eta_{14}$[SOL,DEL] & 0.236 & 0.466 & 0.644 \\
			$\eta_{23}$[VL,BB] & -0.294 & 0.039 & 0.370 \\
			$\eta_{24}$[VL,DEL] & -0.168 & 0.119 & 0.403 \\
			$\eta_{34}$[BB,DEL] & 0.131 & 0.405 & 0.621 \\
			$\rho_{(1)}$[SOL] & -0.109 & 0.007 & 0.174 \\
			$\rho_{(2)}$[VL] & -0.142 & -0.054 & 0.129 \\
			$\rho_{(3)}$[BB] & -0.121 & 0.047 & 0.319 \\
			$\rho_{(4)}$[DEL] & -0.005 & 0.247 & 0.516 \\
			$\gamma$ & -0.060 & 0.019 & 0.136 \\
		\end{tabular}
		\caption{95\% credible intervals and posterior medians of correlations for non-ambulatory data (Section \ref{Posteriors of model parameters})}
		\label{table:DA9}
	\end{table}
	
	\begin{figure}[H]
	\centering
	\includegraphics[width=\textwidth,height=0.95\textheight]{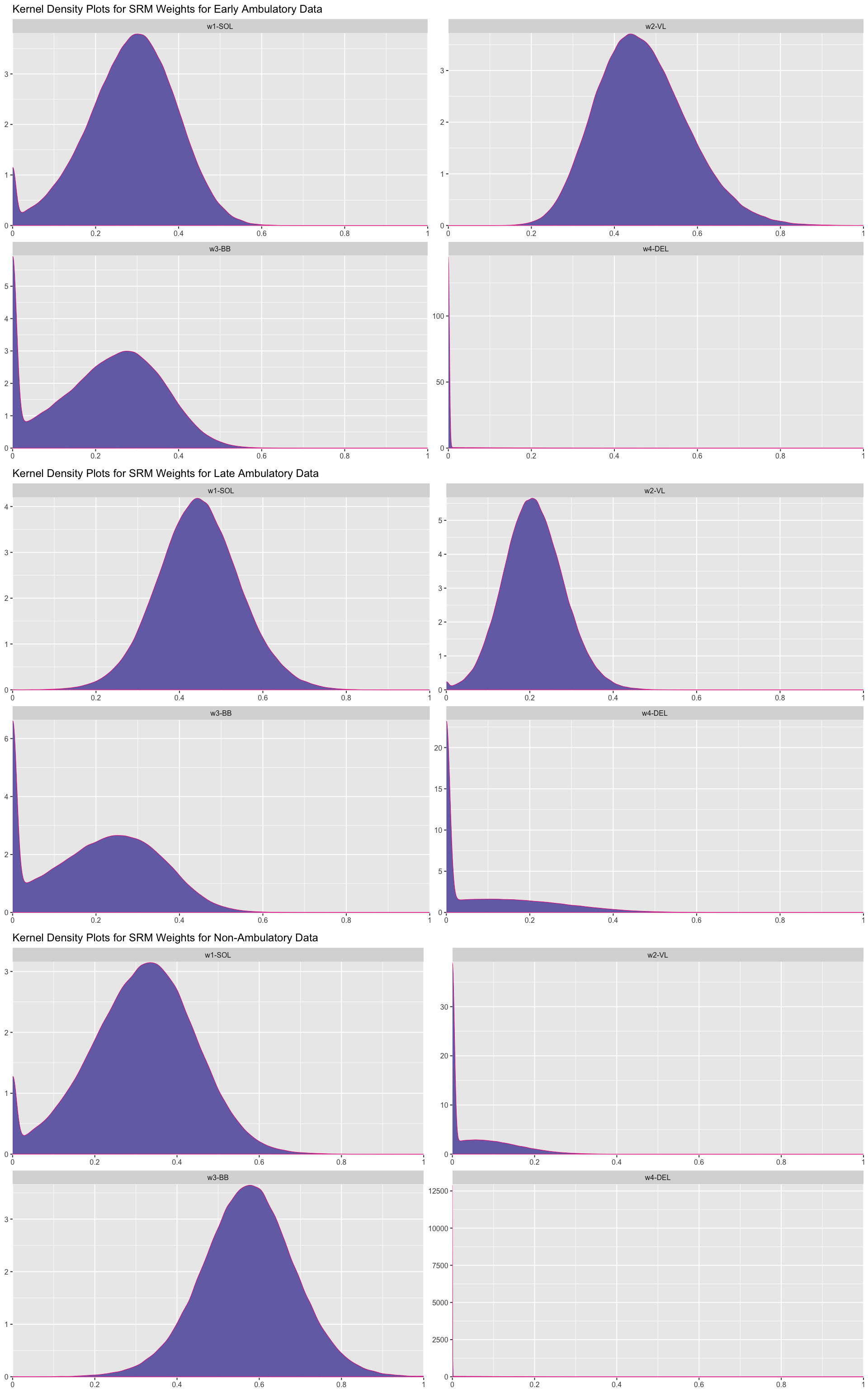}
	\caption{Posterior densities of weights for early to non-ambulatory data (Section \ref{Data analysis - Posteriors of weights and SRM})}
	\label{w_dens.png}
	\end{figure}
	
	\begin{table}[H]
	\centering
		\begin{tabular}{l|ccc}
			Early ambulatory & 2.5th & Point estimate & 97.5th \\
			\hline
			$w_{1}$[SOL] & 0.053 & 0.290 & 0.479 \\
			$w_{2}$[VL] & 0.278 & 0.465 & 0.701 \\
			$w_{3}$[BB] & 0 & 0.245 & 0.452 \\
			$w_{4}$[DEL] & 0 & 0 & 0.200 \\
		\end{tabular}
		\caption{95\% credible intervals and point estimate of weights for early ambulatory data (Section \ref{Data analysis - Posteriors of weights and SRM})}
		\label{table:DA11}
	\end{table}
	\begin{table}[H]
	\centering
		\begin{tabular}{l|ccc}
			Late ambulatory & 2.5th & Point estimate & 97.5th \\
			\hline
			$w_{1}$[SOL] & 0.251 & 0.459 & 0.647 \\
			$w_{2}$[VL] & 0.074 & 0.218 & 0.357 \\
			$w_{3}$[BB] & 0 & 0.235 & 0.459 \\
			$w_{4}$[DEL] & 0 & 0.088 & 0.420 \\
		\end{tabular}
		\caption{95\% credible intervals and point estimate of weights for late ambulatory data (Section \ref{Data analysis - Posteriors of weights and SRM})}
		\label{table:DA12}
	\end{table}
	\begin{table}[H]
	\centering
		\begin{tabular}{l|ccc}
			Non-ambulatory & 2.5th & Point estimate & 97.5th \\
			\hline
			$w_{1}$[SOL] & 0.041 & 0.358 & 0.555 \\
			$w_{2}$[VL] & 0 & 0.051 & 0.252 \\
			$w_{3}$[BB] & 0.344 & 0.591 & 0.796 \\
			$w_{4}$[DEL] & 0 & 0 & 0.242 \\
		\end{tabular}
		\caption{95\% credible intervals and point estimate of weights for non-ambulatory data (Section \ref{Data analysis - Posteriors of weights and SRM})}
		\label{table:DA13}
	\end{table}
	
	
    
    \subsubsection{Visualization of joint posterior distribution of weights (Section \ref{Data analysis - Posteriors of weights and SRM})} \label{Supplement B -- Gemplots}
    To visualize the joint posterior of the weights, we first note that the weights of the four muscles are defined on the unit simplex. As such, each serves as a normalized barycentric coordinate on a tetrahedron, i.e. one can plot the four weights as a point on a tetrahedron by multiplying them with the tetrahedron's vertices \cite{Deaux}. We use the following vertices which define a regular tetrahedron with edge length 1: $(0,0,0)$, $(1,0,0)$, $(1/2,\sqrt{3}/2,0)$, $(1/2,\sqrt{3}/6,\sqrt{6}/3)$. After plotting the weights on the tetrahedron, we use an implementation of a three-dimensional version of the boxplot called \texttt{gemPlot} (\texttt{R} package) \cite{Kruppa} to obtain a visualization of the joint posterior distribution of the weights. For each gemplot in Figure \ref{gem.png} below, the inner yellow polyhedron is called the bag or inner gem and contains 50\% of the points. It is built by extending concepts of halfspace location depths and depth regions from the two-dimensional bagplot \cite{Rousseauw} to the three-dimensional \cite{Kruppa}. The outer red polyhedron is called the loop or outer gem. It is built by first inflating the bag by a factor of 3 (default value \cite{Kruppa} \cite{Gower}) to create the fence, and then the loop is a convex hull of the points in the fence \cite{Gower}. Everything outside of the loop are considered outliers and are shown as small black circles. Non-outliers are inside the loop as blue points. The point of equal weights is indicated as the letter E, which lies outside of the early and non-ambulatory gemplots but inside the late ambulatory gemplot between the inner and outer gems.
    
	\begin{figure}[H]
	\centering
	\includegraphics[width=0.60\textwidth, height=0.95\textheight]{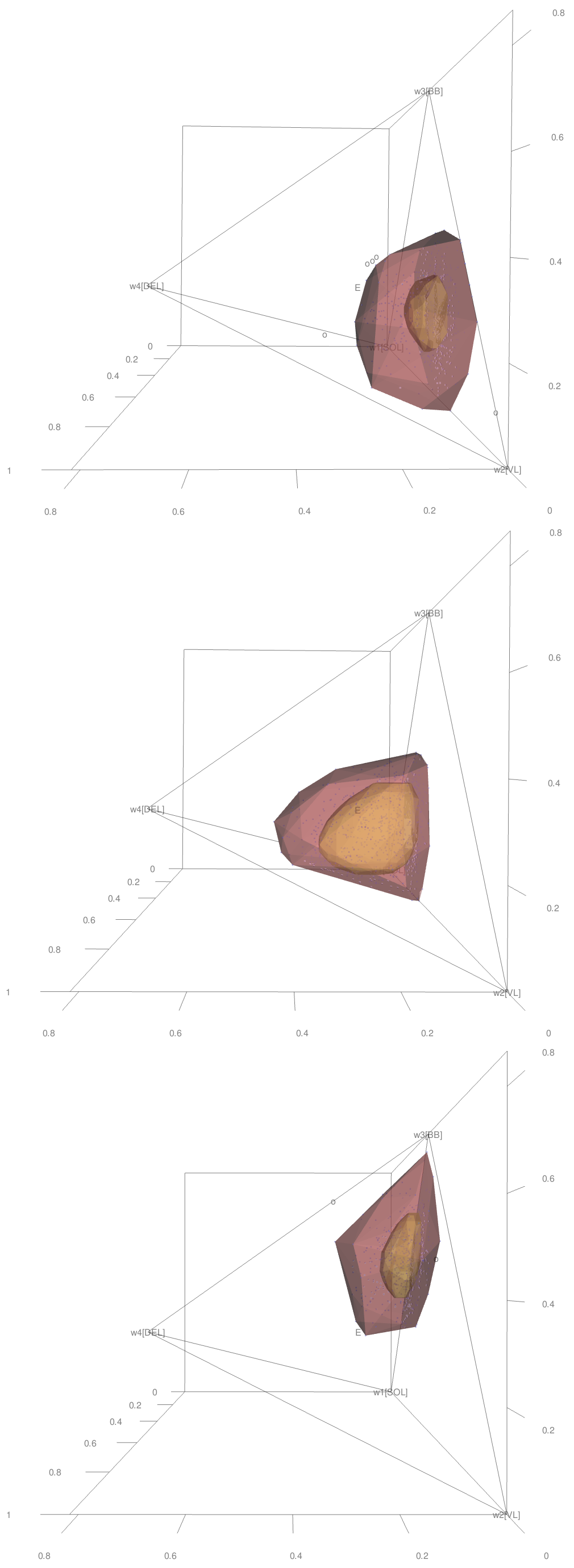}
	\caption{Gemplots of weights for early to non-ambulatory data (top to bottom) (Section \ref{Data analysis - Posteriors of weights and SRM})}
	\label{gem.png}
	\end{figure}
		
\end{document}